{}
{}

\documentclass[12pt,a4paper]{article}

\newif\ifpublic\publicfalse

\usepackage{hyperref}
\usepackage{color}
\usepackage[english]{babel}
\usepackage{amsmath,amssymb}
\usepackage{tensor}
\usepackage{mathabx}
\usepackage{multirow,booktabs}
\usepackage{feynmp-auto}
\usepackage{graphicx}
\usepackage{subcaption}
\usepackage{delimset}

\usepackage[a4paper,text={160mm,247mm},centering]{geometry}
\allowdisplaybreaks[3]
\numberwithin{equation}{section}
\numberwithin{figure}{section}
\newcommand{\nn}{\nonumber}
\newcommand{\nln}{\nonumber\\}

\usepackage[font=small,labelfont=bf,width=0.85\textwidth]{caption}

\expandafter\def\expandafter\bfseries\expandafter{\bfseries\ifmmode\else\boldmath\fi}
\expandafter\def\expandafter\mdseries\expandafter{\mdseries\ifmmode\else\unboldmath\fi}
\expandafter\def\expandafter\normalfont\expandafter{\normalfont\ifmmode\else\unboldmath\fi}

\RequirePackage{verbatim}

\makeatletter
\newwrite\bibinl@out

\makeatother

\let\barefrac=\frac
\renewcommand{\frac}[2]{\mathinner{\barefrac{#1}{#2}}}

\let\baresqrt=\sqrt
\makeatletter
\renewcommand{\sqrt}{\@ifnextchar[\@sqrt@space@a\@sqrt@space@b}
\def\@sqrt@space@a[#1]#2{\mathinner{\mathchoice{\mkern-3mu}{\mkern-3mu}{}{}\baresqrt[#1]{#2}}}
\def\@sqrt@space@b#1{\mathinner{\mathchoice{\mkern-3mu}{\mkern-3mu}{}{}\baresqrt{#1}}}
\makeatother

\makeatletter
\let\per@dot@old=\.
\def\.{\ifmmode\def\per@dot@sel{\mkern3mu}\else\def\per@dot@sel{\per@dot@old}\fi\per@dot@sel}
\makeatother

\let\barefootnote=\footnote
\renewcommand{\footnote}[1]{\barefootnote{#1\vspace{3pt}}}

\newcommand{\sfrac}[2]{{\textstyle\frac{#1}{#2}}}
\newcommand{\half}{\sfrac{1}{2}}

\newcommand{\vfrac}[2]{\ifmmode\mathinner{\textstyle^{#1}\!/\!_{#2}}\else$^{#1}\!/\!_{#2}$\fi}

\newcommand{\master}{\widehat \delta}
\newcommand{\eunit}{\mathinner{\mathrm{e}}}
\newcommand{\la}{\lambda}
\newcommand{\s}{\sigma}

\DeclareMathOperator{\tr}{tr}

\newcommand{\ft}[2]{{\textstyle\frac{#1}{#2}}}

\newcommand{\dx}{{\dot x}}
\newcommand{\ddx}{{\ddot x}}

\newcommand*{\diff}{{\mathrm d}} 
\newcommand{\plexp}{\overleftarrow{\mathrm{P}} \hspace*{-1.4mm} \exp}
\newcommand{\prexp}{\overrightarrow{\mathrm{P}} \hspace*{-1.4mm} \exp}

\newcommand{\Order}{\mathcal{O}}

\newcommand{\alg}[1]{\mathfrak{#1}}
\newcommand{\grp}[1]{\mathrm{#1}}

\def\<{\big\langle}
\def\>{\big\rangle}
\newcommand{\unit}{\mathbf{1}}                              

\newcommand{\geom}[1]{\mathrm{#1}}

\newcommand{\AdS}{\geom{AdS}}

\newcommand{\Reals}{\mathbb{R}}

\providecommand{\href}[2]{#2}

\providecommand{\hypersetup}[1]{}
\providecommand{\texorpdfstring}[2]{#1}
\hypersetup{plainpages=false}
\hypersetup{pdfpagemode=UseNone}
\hypersetup{bookmarksnumbered=true}
\hypersetup{pdfstartview=FitH}
\hypersetup{colorlinks=false}
\hypersetup{citebordercolor={1 1 1}}
\hypersetup{urlbordercolor={1 1 1}}
\hypersetup{linkbordercolor={1 1 1}}

\makeatletter
\let\@keywords\@empty
\let\@subject\@empty
\providecommand{\keywords}[1]{\gdef\@keywords{#1}}
\providecommand{\subject}[1]{\gdef\@subject{#1}}
\def\thetitle{\@title}
\def\theauthor{\@author}
\def\thesubject{\@subject}
\def\thedate{\@date}
\def\thekeywords{\@keywords}
\makeatother
\AtBeginDocument{
\hypersetup{pdftitle={\thetitle}}
\hypersetup{pdfauthor={\theauthor}}
\hypersetup{pdfsubject={\thesubject}}
\hypersetup{pdfkeywords={\thekeywords}}}

\newif\ifshownote
\shownotetrue

\ifpublic\shownotefalse\fi

\ifshownote

\ifpdf\else\RequirePackage[active]{srcltx}\fi
\RequirePackage{xcolor}
\RequirePackage{ulem}

\newcommand{\remark}[2][]{{\normalfont\sffamily\hspace{1ex}
\def\emph{\textsl}\def\textbullet{$\bullet$}
\def\tmparga{#1}
\def\tmpargb{HM}\ifx\tmparga\tmpargb\color[rgb]{0.7,0,0}\fi
\def\tmpargb{}\ifx\tmparga\tmpargb\color{red}\fi
\def\tmpargb{}\ifx\tmparga\tmpargb\else \textbf{#1:}\fi
#2\hspace{1ex}}}

\else
\newcommand{\remark}[2][]{\ignorespaces}

\fi

\def\[{\begin{equation}}
\def\]{\end{equation}}

\title{Wilson Loops and Integrability}
\author{Hagen M{\"u}nkler}

\begin{document}
\thispagestyle{empty}


\begin{center}
{\Large\bfseries Wilson Loops and Integrability \par}
\vspace{12mm}

\begingroup\scshape\large 
Hagen M{\"u}nkler
\endgroup
\vspace{5mm}

\textit{Institut f\"ur Theoretische Physik, Eidgen{\"o}ssische Technische Hochschule Z{\"u}rich, \phantom{$^\S$}\\
Wolfgang-Pauli-Strasse 27, CH-8093 Z{\"u}rich, Switzerland. } \\[0.1cm]
\texttt{ \small{ muenkler@itp.phys.ethz.ch }} \\ \vspace{5mm}

\vspace{5mm}

\textbf{Abstract}\vspace{5mm}\par
\begin{minipage}{14.7cm}
These notes provide an introduction toward Wilson loops in $\mathcal{N} \!=4$ supersymmetric
Yang-Mills theory with a focus toward their integrability properties. In addition to a brief 
discussion of exact results for the circular Wilson loop and the cusp anomalous dimension, the 
notes focus on non-local symmetries, utilizing the integrability of the minimal surface problem
that appears at strong coupling. 
This work is based on lectures given at the Young Researchers Integrability School and Workshop 2018. 
To appear in a special issue of J.\ Phys.\ A.
\end{minipage}\par
\end{center}

\setcounter{tocdepth}{2}
\hrule height 0.75pt
\tableofcontents
\vspace{0.8cm}
\hrule height 0.75pt
\vspace{1cm}

\section{Introduction}
\label{sec:introduction}

The below review is based on lectures given at the 2018 edition of the Young Researchers 
Integrability School and Workshop and gives an introduction to Wilson loops with a 
focus toward the Maldacena--Wilson loop in $\mathcal{N}\! = 4$ supersymmetric 
Yang--Mills theory. 

The Wilson loop is a non-local observable which can be considered in any gauge theory 
and is important both for the study of confinement as well as for the infrared singularities 
of scattering amplitudes. In $\mathcal{N}\!=4$ supersymmetric Yang--Mills theory, one often 
considers the Maldacena--Wilson loop, which is a generalization of the Wilson loop specific to 
this theory, where it is perhaps an even more central observable than the Wilson loop 
is in other theories. For example, it appears to be dual to scattering amplitudes for certain 
configurations whereas other configurations allow for exact calculations, which can be employed 
to test the AdS/CFT correspondence. 

The discussion of Wilson 
loops in generic gauge theories is restricted to their renormalization properties as 
well as the relation to the quark-antiquark potential. 
For the Maldacena--Wilson loop, we discuss --- in addition to the above-mentioned results ---
its strong-coupling description in terms of minimal surfaces in $\mathrm{AdS}_5$ in detail. 
This will lay the foundation for the discussion of hidden symmetries of the Maldacena--Wilson 
loop which concludes these lecture notes. There we make use of the fact that 
the minimal surface is described by an integrable, classical theory in order 
to extract Yangian symmetries. Related algebraic structures are discussed in 
the review on One-point functions in AdS/dCFT \cite{deLeeuw:2019usb} to appear 
in the same special issue of J.\ Phys.\ A.

\section{Wilson Loops in Yang--Mills Theories}
\label{sec:WL}

We begin by discussing Wilson loops in generic non-Abelian Yang--Mills theories, where they 
were first considered by Wilson \cite{Wilson:1974sk} in the study of quark confinement using 
lattice methods. 
Here, we follow \cite{Dorn:1986dt} and take a geometric approach, which introduces the 
Wilson loop as the parallel transport in the gauge theory. This underlines the connection 
to the monodromy which we will employ in the discussion of integrability for minimal surfaces. 

Here, we will assume a gauge theory with Yang--Mills coupling constant $g$, fundamental fermion 
fields $\psi$ and gauge field $A_\mu$, which we expand as 
\begin{align}
A_\mu &= A_\mu ^a \, t^a \, , & 
\tr ( t^a \, t^b ) = \half \delta ^{ab} , 
\end{align} 
in terms of the generators $t^a$ of the Lie algebra of the gauge group. Moreover, we 
have the covariant derivative and field strength
\begin{align}
D_\mu \psi &= \partial_\mu \psi - i A_\mu \psi \, , &
F_{\mu \nu} &= \partial_\mu A_\nu - \partial_\nu A_\mu - i \left[A_\mu , A_\nu \right] .
\end{align}
Note that we cannot compare the values of the field $\psi$ at two points $x,y \in \Reals^{(1,3)}$ 
directly, since they do not transform in the same way under gauge transformations,
\begin{align}
\psi(x) \; \; &\mapsto \; \; U(x) \psi(x) \, , &
\psi(y) \; \; &\mapsto \; \; U(y) \psi(y) \, ,
\end{align}
where $U(x)$ is an element of the gauge group. 
One encounters the same problem for tangent vectors at different points of a manifold
and we approach it in the same way by introducing the notion of parallel transport along a curve. 
In the context of gauge theories, the parallel transport is known as the Wilson line
and can be introduced by requiring that it be covariantly constant along a path 
connecting the points $x$ and $y$. 
More explicitly, consider a curve $\gamma$ with parametrization $x(\sigma)$ from $y$ to $x$ 
and construct the Wilson line $V_\gamma(x(\sigma), y)$ from the differential equation 
$\dx ^\mu D_\mu V_\gamma = 0$, or more explicitly
\begin{align}
\frac{\diff}{\diff \sigma} \, V_\gamma(x(\sigma), y) 
	= i \dx^\mu (\sigma) A_\mu (x(\sigma)) \, V_\gamma(x(\sigma), y) .
\label{def:WilsonLine}
\end{align}
Together with the initial condition $V_\gamma(y, y)= \unit$, this equation determines the 
Wilson line completely as can be seen from the uniqueness theorem for ordinary differential 
equations. The Wilson line thus inherits the usual properties of the parallel transport. In the 
case of a concatenation $\gamma_2 \ast \gamma_1$ of two curves $\gamma_1$ and $\gamma_2$, for example,
we have 
\begin{align}
V_{\gamma_2 \ast \gamma_1}(z,x) = V_{\gamma_2}(z,y) \, V_{\gamma_1}(y,x) \, , 
\end{align}
for some point $z$ located along $\gamma_2$ and $y$ denoting the connecting point of the two curves. 
The proof of the above statement is a simple consequence of the uniqueness theorem for ordinary differential 
equations: It is easy to see that the right-hand side satisfies the defining equation \eqref{def:WilsonLine} 
for the Wilson line over $\gamma_2 \ast \gamma_1$ for $z$ located along $\gamma_2$ and the factor of 
$V_{\gamma_1}(y,x)$ ensures that it depends on $\sigma$ in a smooth way as long as the contour is smooth 
as well.  

The behaviour of the Wilson line under gauge transformations 
\begin{align*}
A_\mu \mapsto A_\mu ^\prime = U \left( A_\mu + i \partial_\mu \right) U^{-1}
\end{align*}
can be established in the same way and one finds that the Wilson line transforms as
\begin{align}
V_\gamma (x,y) \; \; &\mapsto \; \; V_\gamma ^\prime (x,y) = U(x) V_\gamma(x,y) U^{-1}(y) .
\end{align}

We have thus reached our goal to be able to compare the field $\psi$ at different points, 
since $\psi(x)$ and $V_\gamma (x,y) \psi(y)$ transform in the same way under gauge transformations. 
Moreover, if we have e.g.\ scalar fields $\Phi$ in the adjoint representation as in $\mathcal{N}\!=4$ supersymmetric Yang--Mills theory, we can construct non-local gauge invariant operators such as
\begin{align*}
\tr \left( \Phi (x) V_\gamma (x,y) \Phi(y) V_{\gamma} (y,x) \right) . 
\end{align*}
Another possibility is to consider a closed curve $\gamma$, for which the Wilson line transforms as
\begin{align}
V_\gamma (x,x) \; \; \mapsto \; \; U(x) V_\gamma (x,x) U(x)^{-1} .  
\label{W_gauge}
\end{align}
This shows that all eigenvalues of $V_\gamma (x,x)$ are gauge-invariant. The Wilson loop is a 
specific combination of these gauge-invariant quantities, the trace
\begin{align}
\mathrm{W}(\gamma) = \frac{1}{N} \, \tr \left(  V_\gamma (x,x) \right) .
\end{align}
Here, $N$ is the dimension of the fundamental representation of the gauge group, which we will 
take to be $\grp{SU}(N)$ from now on. The normalization factor ensures that the trivial loop 
over a constant curve gives $\mathrm{W}(\gamma)=1$. One can also consider other 
representations of the gauge group and construct the Wilson loop there; this is related to 
considering other combinations of the eigenvalues. 
Here, we focus on the Wilson loop in the fundamental representation, which we have obtained
by considering the gauge transformation properties of a fermion field transforming in 
the fundamental representation of the gauge group. 

The Wilson loop is typically written in a different form, which we obtain by rewriting 
the defining equation \eqref{def:WilsonLine} as an integral equation,
\begin{align}
V_\gamma (x(\sigma),y) = \unit + i \int \limits _0 ^\sigma \diff \sigma_1 \, 
	\dx_1 ^\mu A_\mu (x_1) V_\gamma (x_1,y) \, ,
\end{align} 
where we have abbreviated $x(\sigma_1) = x_1$. By iteratively plugging this recursion into itself, 
we obtain the formal solution 
\begin{align}
V_\gamma (x(\sigma),y) = \plexp \left( i \int _0 ^\sigma \diff \sigma_1 \, 
	\dx_1 ^\mu A_\mu (x_1) \right) , 
\end{align}
where the arrow indicates that in the expansion of the path-ordered exponential, greater values of $\sigma$ are ordered to the left. For the Wilson loop we thus have the expression
\begin{align}
\mathrm{W}(\gamma) = \frac{1}{N} \, \tr 
	\plexp \left( i \int _\gamma \diff \sigma \, \dx^\mu A_\mu (x) \right) ,
\end{align}
which we will use to carry out calculations in perturbation theory. The reader should note that 
in an expectation value, the time-ordering does not override the path-ordering since the 
respective orderings concern different objects. The time-ordering affects the coefficients 
$A_\mu ^a$ whereas the path-ordering refers to the generators $t^a$. 

\subsection{The Quark-antiquark potential}

Above, we have introduced the Wilson loop from a mathematical perspective. Physically, we 
can interpret it as describing the insertion of a heavy external quark into the theory. 
For a brief motivation of this interpretation, we turn to pure electrodynamics, i.e.\ 
pure Yang--Mills theory with gauge group $\grp{U}(1)$. 
Let us first recall the action of the electromagnetic field in the presence of electrons, which 
is given by
\begin{align}
S = -  \sum \limits  _{\text{part.}} m \int \diff s 
	- \sum \limits _{\text{part.}} e \int A_\mu \, \diff x^\mu 
	- \int \diff^4 x F_{\mu \nu} \, F^{\mu \nu} .
\label{eqn:QED-Action}	
\end{align}
Here, the first term describes the action of a free particle, which is simply given by the length 
of its world-line. The second term describes the interaction between the electrons and the 
electromagnetic field, whereas the third term describes the electromagnetic 
field itself.  

We thus see that the expectation value of the Wilson loop, 
\begin{align*}
\left \langle W(\gamma) \right \rangle = 
	\frac{1}{Z_0} \int \left[ \diff A \right] \, \exp \left( i S_{\text{YM}} 
	+ i e \int _\gamma \diff x^\mu A_\mu \right) , 
\end{align*} 
describes the insertion of an external charged particle into the theory. Note that here the 
world-line of the particle is fixed by the contour of the Wilson loop and does not react to 
the electromagnetic field. Correspondingly, the action of the free particle does not need to 
appear, since the contour of the particle does not vary. 

Let us now consider a specific contour, a rectangle with 
side-length $T$ in the time direction and spatial extent $R$: 
\begin{equation*}
\includegraphics{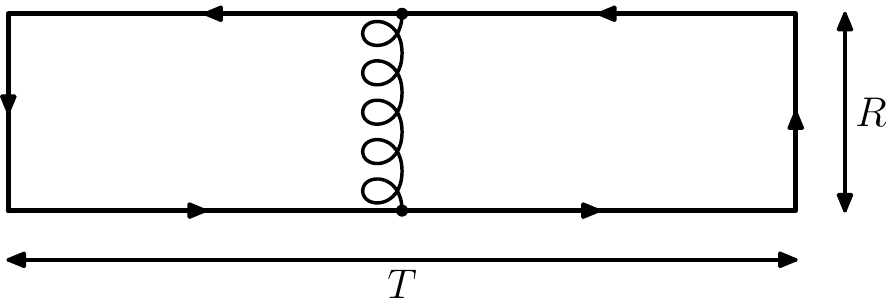}
\end{equation*}
Here, we consider $T$ to be 
much larger than $R$, such that we can neglect the two space-like lines closing the rectangle. 
The Wilson loop over this contour thus describes the insertion of two heavy, static particles 
at a spatial distance $R$ from each other. Since the contour is oriented in the positive time 
direction for the one particle and the negative time direction for the other particle, we view 
them as a particle-antiparticle pair.    

Recall now the quantum-mechanical derivation of the path integral as describing the transition 
amplitude from an initial state at time ${-T/2}$ into a final state at time ${T/2}$. This 
amplitude is described by a superposition of the propagation of energy eigenstates with phase 
factors $\exp(-i E_n T)$. After a Wick rotation to Euclidean time, the ground-state energy 
will dominate this superposition for asymptotically large times $T$. Taking also the normalization 
into account and recalling that we are considering a charged particle-antiparticle pair at 
spatial distance $R$, we find that for large Euclidean times $T$ the expectation value of the 
Wilson loop is given by
\begin{align}
\left \langle \mathrm{W}(\gamma_{R,T} ) \right \rangle \simeq e^{- T \, V(R)} ,  
\end{align}
where $V(R)$ describes the potential between the particle and antiparticle. 
The above result also holds in non-Abelian Yang--Mills theories, for a derivation in this case the
reader is referred to the literature on lattice gauge theory, e.g.\ reference \cite{Smit:2002ug}.  
The calculation of the expectation value of the Wilson loop is hence crucial in the study of 
confinement, which is the problem Wilson originally addressed in reference \cite{Wilson:1974sk}.
We note that in a conformal field theory scale invariance requires that the expectation value is 
of the form $e^{-T/R}$, such that we obtain the Coulomb potential.

An interesting application of the above result is the derivation of the Coulomb potential
from pure quantum electrodynamics. Since the theory is free, we can calculate the expectation value 
of the Wilson loop exactly. First, one may show that the expectation value for a generic contour can 
be written as 
\begin{align*}
\left \langle W(\gamma) \right \rangle = \exp \brk*{
	- \frac{e^2}{8 \pi ^2} \int \diff \s_1  \diff \s_2 
	\frac{\dx_1 \dx_2}{(x_1 - x_2)^2} } .
\end{align*}
Carefully considering the limit $T \gg R$ then allows to derive the Coulomb potential from 
the Wilson loop. 

\subsection{Divergences and Renormalization}
In the perturbative calculation of the expectation value of the Wilson loop, we encounter divergences 
which need to be renormalized. Below, we discuss these divergences for the one-loop approximation 
where they were first observed \cite{Polyakov:1979ca}. The simple calculations performed there are 
sufficient to demonstrate the origin of the divergences and explain their renormalization. For a 
proof of the renormalizability of the expectation value of the Wilson loop, the reader is referred 
to the original literature \cite{Dotsenko:1980wb,Brandt:1981kf}. 
At the one-loop level, the expectation value is given by 
\begin{align}
\left \langle \mathrm{W}(\gamma) \right \rangle = 
	1 - \frac{g^2 (N^2-1)}{16 \pi^2 N} \int \limits _0 ^L \diff \sigma_1 \, 
	\diff \sigma_2 \, 
	\frac{\dx_1 \dx_2 }{(x_1-x_2)^2} + \Order ( g^4) \, . 
\end{align}
Here and below we omit writing out the dot product explicitly and 
we have inserted the gauge field propagator in Feynman gauge,  
\begin{align}
\left \langle A_\mu ^a (x_1) A_\nu ^b (x_2) \right \rangle 
	= \frac{g^2}{4 \pi^2} \, \frac{\eta_{\mu \nu} \, \delta ^{ab}}{(x_1-x_2)^2} , 
\label{eqn:1-loop-exp-value}	
\end{align}
as well as the normalization $\tr (t^a t^b ) = \half \delta^{ab}$. Note that the path-ordering 
was not relevant at this order in perturbation theory. Moreover, we will restrict 
the parametrization of the curve to satisfy $\dx^2 = 1$, such that the parameter $\sigma$ 
corresponds to the arc-length. The use of such a parametrization is indicated above by 
writing the explicit integration boundaries 0 and $L$, even though the form given there is 
still reparametrization invariant. 

The integrand is divergent when the two points $x_1 = x(\sigma_1)$ and $x_2 = x(\sigma_2)$ 
approach each other. Here, we assume that the curve does not intersect itself, such that this 
happens when $\sigma_1$ and $\sigma_2$ approach each other or at the end point of the closed 
curve. Let us first consider the divergence coming from $\sigma_1 \sim \sigma_2$. Here, we employ 
a cut-off regularization, which modifies the position space propagator as 
\begin{align*}
\frac{1}{x^2} \; \to \; \frac{1}{x^2 + a^2} \, ,
\end{align*}
and we consider the limit $a \to 0$. When the two parameters are close to each other, we may 
calculate the divergent part of the one-loop expectation value \eqref{eqn:1-loop-exp-value}
as follows:
\begin{align}
&\int \limits _0 ^L \diff s \int \limits _{-s} ^{L-s} \diff t \, 
	\frac{\dx(s) \dx(s+t)}{[x(s+t)-x(s)]^2 + a^2 } 
	\simeq \int \limits _0 ^L \diff s \int \limits _{-s} ^{L-s} \diff t \, 
	\frac{1}{t^2 + a^2} \, + ( \text{finite}) \nln
&= 2 \int \limits _0 ^{L/a} \diff \sigma \arctan(\sigma) + ( \text{finite})
	= 2  \frac{L}{a} \arctan \brk*{ \frac{L}{a} } 
	- \ln \brk*{ 1 + \frac{L^2}{a^2} } + ( \text{finite}) \nln
&= \frac{\pi L}{a} - 2 \ln \brk*{ \frac{L}{a} }  + ( \text{finite}) .
\label{Linear_Div_1}	
\end{align}
In the first step, we have neglected all higher-order corrections in $t$ which are due 
to the curvature of the contour. Indeed, they do not contribute to the divergent part of 
the result, which we have effectively calculated for a straight line of length $L$ above.  
This calculation, however, overlooks that $x(L-\sigma)$ and $x(\sigma)$ are also close to 
each other for small $\sigma$, since we are considering a closed curve. 

The divergent contribution arising from integrating close to the starting and end point 
of the contour can be captured in the expression (we are using a periodic parametrization) 
\begin{align}
& \int \limits _{-L/4} ^0 \diff \sigma_1 \int \limits _0 ^{L/4} \diff \sigma_2 
	\frac{\dx_1 \dx_2}{(x_2 - x_1)^2 + a^2} 
= \int \limits _0 ^{L/4} \diff \sigma_1 \diff \sigma_2 
	\frac{1}{(\s_1 + \s_2)^2 + a^2} + (\text{finite})  \nln
&= \int \limits _0 ^{L/4a} \diff \sigma \brk*{ \arctan \brk*{ \sfrac{L}{4a} + \sigma} 
	- \arctan (\sigma) } + (\text{finite}) 	
	= \ln \brk*{ \frac{L}{a} }  + ( \text{finite}) .	
\label{Linear_Div_2}		
\end{align}
Note that choosing the integration boundaries to be $\pm \sfrac{L}{4}$ was not relevant for 
the calculation of the divergent part, but only made sure that the points $x(\sigma)$ do not 
approach each other for non-zero values of $\sigma$. The contribution discussed above appears 
twice in the calculation of the one-loop expectation value, since we need to take both orderings 
of $\sigma_1$ and $\sigma_2$ into account. The logarithmic divergence thus cancels between the 
terms \eqref{Linear_Div_1} and \eqref{Linear_Div_2}, such that we are left with the linear 
divergence in the case of a smooth curve,   
\begin{align}
\int \limits _0 ^L \diff \sigma_1 \, \diff \sigma_2 \, 
	\frac{\dx_1 \dx_2 }{(x_1-x_2)^2 + a^2 } 
	= \frac{\pi L}{a}  + ( \text{finite}) . 
\label{Linear_Div}	 
\end{align}
This linear divergence appears in all orders of perturbation theory and 
$G(\gamma) = \left \langle \mathrm{W}(\gamma) \right \rangle$ 
can be renormalized \cite{Dotsenko:1980wb} as
\begin{align}
G_\mathrm{ren}(\gamma) 
	= e^{-\delta m \, L(\gamma) } \, G (\gamma)  \, ,
\end{align}
which --- remembering the action \eqref{eqn:QED-Action} for an electron in an electromagnetic 
field --- we may interpret as a mass renormalization of the external particle described 
by the Wilson loop. 

In the case of an open end or a cusp, the Wilson loop has additional divergences. 
Note first that our calculations above show that an open Wilson line has logarithmic end-point divergences, since 
the calculation leading to equation \eqref{Linear_Div_1} is still correct, but the cancellation with 
the contribution \eqref{Linear_Div_2} no longer appears. In the case of a cusp (located at $x(0)$ 
for convenience), both contributions are present, but the calculation of the second term needs to 
be adapted to include the cusp angle and the cancellation of the logarithmic contributions 
no longer takes place. 

For the discussion of the cusp divergences, we will switch to dimensional regularization, which 
is more commonly used than the cut-off regularization we discussed above. In dimensional 
regularization, the momentum space propagators are unaltered, but the Fourier transformation 
is carried out in $\mathrm{D} =4 - 2 \epsilon$ dimensions. The two-point functions are then modified as
\begin{align*}
\frac{1}{x^2} \; \to \; \frac{1}{(x^2)^{1-\epsilon}} , 
\end{align*}
cf.\ e.g.\ reference \cite{Erickson:2000af} for more details. Now, the logarithmic divergences
associated to the cusps appear as poles in the expansion in $\epsilon$. In order to calculate the
cusp anomalous dimension at the one-loop order, we consider the following diagrams:
\vspace*{3mm}
\begin{equation*}
\includegraphics{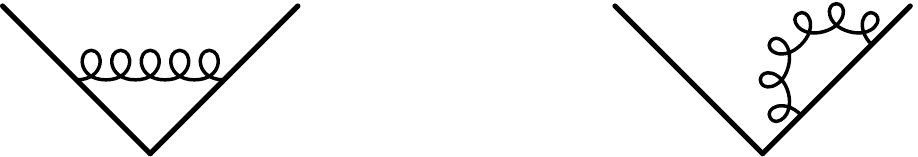}
\end{equation*}
Clearly, the angle-dependence is contained in the first diagram, whereas the second diagram 
can only contribute a constant term. The relevant integral for the one-loop calculation of 
the cusp divergence thus comes from considering one integration along each of the intersecting
lines. Denoting the angle between the two lines by $\phi$, we get
\begin{align*}
F(\phi) = \int \limits _0 ^L 
	\frac{\diff \sigma_1 \, \diff \sigma_2 \, \cos(\phi)}{\left[\sigma_1 ^2 + \sigma_2 ^2 
	+ 2 \sigma_1 \sigma_2 \cos(\phi) \right] ^{1 - \epsilon} }
	&= \int \limits _0 ^L \frac{\diff \ell}{\ell ^{1 - 2 \epsilon}}
	\int \limits _0 ^1  \frac{\diff z \, \cos(\phi)}{z^2 + \bar{z}^2 
	+ 2 z \bar{z} \cos(\phi)} + \Order (\epsilon^0) \\
	&= \frac{\phi \cot(\phi)}{2 \epsilon} + \Order (\epsilon^0) .
\end{align*}
Here, we have used the substitution $\sigma_1 = \ell z$, $\sigma_2 = \ell \bar{z} = \ell (1-z)$;
the divergence is then captured in the scale integral over $\ell$. We have seen above that also 
the second diagram leads to a log-divergence and our above result does hence not 
capture the whole divergence. Note however, that the log-divergence of this diagram has to cancel 
with the one obtained from the first diagram with the intersection angle set to zero.  
The divergence of the cusped Wilson loop is thus given by
\begin{align}
W ( < ) \sim \brk!{ F(\phi) - F(0) }
	= \frac{\phi \cot(\phi) - 1}{2 \epsilon} + \Order (\epsilon^0) .
\end{align}
The cusp divergence is renormalized multiplicatively through a $\phi$-dependent $Z$-factor
\begin{align}
\left \langle W_{\text{ren}} (\gamma) \right \rangle 
	= Z(\phi) \, \left \langle W (\gamma) \right \rangle \, , 
\end{align}
with the same $Z(\phi)$ for all curves $\gamma$ with the same cusp angle $\phi$. 
Here, we have omitted the dependence on the regulator which the quantities appearing on the 
right-hand side have. The renormalization of a cusped Wilson loop appears in addition to the 
usual renormalizations to be considered in the gauge theory and the renormalization group 
equation for the cusped Wilson loop is given by
\begin{align}
\brk*{ \mu \frac{\partial}{\partial \mu} + \beta (g_R) \frac{\partial}{\partial g_R} 
	+ \Gamma_{\text{cusp}} (\phi , g_R) } \left \langle W_{\text{ren}} (\gamma) \right \rangle = 0 \, ,  
\end{align}
where $\beta (g_R)$ describes the dependence of the coupling constant on the renormalization scale 
and $\Gamma_{\text{cusp}} (\phi , g_R)$ is known as the cusp anomalous dimension. It is presently 
known up to three loops \cite{Korchemsky:1987wg,Grozin:2015kna} in QCD and up to four loops in  
$\mathcal{N}\!=4$ supersymmetric Yang--Mils theory \cite{Makeenko:2006ds,Correa:2012nk,Henn:2013wfa}. 

Similar divergences appear for Wilson loops with self-intersections. In this case, however, the 
renormalization mixes the original operator with correlators of Wilson loops taken over the same 
contour but with different orderings around the intersection point. For the simplest example of 
a single intersection, the renormalization mixes between the following contours:
\begin{equation*}
\includegraphics{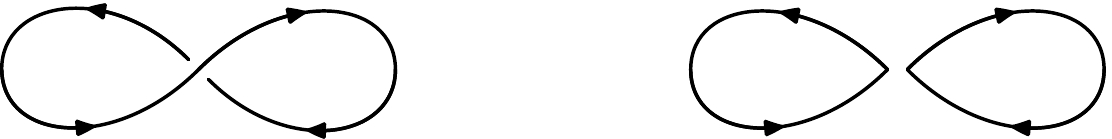}
\end{equation*}

The anomalous dimension then becomes a matrix in the space of the contours which are mixed by 
the renormalization and is known as the cross or soft anomalous dimension. It is an important 
quantity in the description of the infrared divergences 
of scattering amplitudes, see e.g. reference \cite{Magnea:2014vha} for a review or 
\cite{White:2015wha} for a pedagogical introduction to the modern methods used in these calculations. 
Intuitively, we can understand the connection between the UV divergences of Wilson loops and the IR 
divergences of scattering amplitudes as follows: The Wilson line describes an external quark 
following its classical straight-line trajectory. For the emission of soft gluons 
of zero momentum, this trajectory is a valid approximation and the Wilson loop accounts for the 
acquired phase factor.  

\section{The Maldacena--Wilson Loop in \texorpdfstring{$\mathcal{N}\!=4$}{N=4} SYM}
\label{sec:MWL}

The Maldacena--Wilson loop is a generalization of the Wilson loop which is specific to 
$\mathcal{N}\!=4$ supersymmetric Yang--Mills theory. Maldacena's original derivation originated from 
considering $(N+1)$ three-branes and separating one of them from the others. He thus studied the Higgs 
mechanism for the symmetry breaking  $\grp{U}(N+1) \to \grp{U}(N) \times \grp{U}(1)$. For an account of 
this approach, the reader is referred to the original papers \cite{Maldacena:1998im,Rey:1998ik} or the textbook 
\cite{Nastase:2015wjb}. Here, we consider the dimensional reduction of light-like Wilson loops 
in ten-dimensional $\mathcal{N}\!=1$ supersymmetric Yang--Mills theory, which facilitates the 
discussion of the local supersymmetry of the operator.   

Let us shortly recall some basics of the dimensional reduction. The ten-dimensional theory contains 
the gauge field $A_m$ and a ten-dimensional Majorana--Weyl spinor $\Psi$. Both fields take values in  
the Lie algebra $\alg{su}(N)$ and the action is of the form 
\begin{align}
S \; \propto \; \int \diff ^{10} x \, 
	\tr \left( - \half \, F_{mn} \, F^{mn} + i \bar{\Psi} \, \Gamma^m D_m \, \Psi \right) . 
\label{10daction}	
\end{align}
Here, the matrices $\Gamma_m$ are ten-dimensional Dirac matrices, which satisfy a ten-dimensional 
Clifford algebra. The action is invariant under the supersymmetry transformations
\begin{align}
\delta _\xi A_m &= i \, \bar{\xi} \, \Gamma_m \, \Psi \, , & 
\delta _\xi \Psi &= - \ft{1}{4} F_{mn} \, \brk[s]*{\Gamma ^m , \Gamma ^n} \, \xi \, ,
\label{10dSUSY}
\end{align}
where the supersymmetry parameter $\xi$ is a constant, ten-dimensional Majorana--Weyl spinor. 
The dimensional reduction to four dimensions is obtained by demanding that the fields only depend 
on the coordinates $x^\mu$ of $\mathbb{R}^{(1,3)} \subset \mathbb{R}^{(1,9)}$. This implies that 
the last six components of the gauge field $A_m$ do not transform as gauge fields any more, but 
simply in the adjoint representation 
\begin{align}
\Phi_I = A_{I+3}  \; \; & \mapsto \; \; U(x) \Phi_I U(x) ^\dag .
\end{align}
Moreover, from the four-dimensional viewpoint, i.e.\ with respect to the Lorentz group in 
$\mathbb{R}^{(1,3)}$, they are scalar fields. The ten-dimensional spinor field $\Psi$ can 
also be decomposed into a set of four-dimensional spinor fields but that is not our concern 
here. The action of the four-dimensional theory inherits the invariance under the supersymmetry 
transformations \eqref{10dSUSY} from the ten-dimensional theory, which appears as 
$\mathcal{N}\!=4$ supersymmetry after decomposing the ten-dimensional spinor into four-dimensional
spinors as for the fermion fields. Due to the additional presence of conformal invariance, the 
symmetry algebra of the four-dimensional theory is lifted to the superconformal algebra 
$\alg{psu}(2 , 2 \vert 4)$.  

The Wilson loop in ten-dimensional $\mathcal{N}\!=1$ supersymmetric Yang--Mills theory is given by
\begin{align}
\mathrm{W}(\gamma) = \frac{1}{N} \, \tr \plexp 
	\left( i \int _\gamma A_m \, \diff x^m \right) .
\end{align}
The linear divergence we discussed above is absent for a light-like contour, but new divergences 
appear in this case \cite{Korchemskaya:1992je}. These new divergences, however, will not appear 
in the four-dimensional theory for contours which are not light-like in four dimensions, 
as we assume in the following. Let us now consider the supersymmetry variation 
of the Wilson loop. 
Using the field variation \eqref{10dSUSY}, we find 
\begin{align}
\delta_\xi \mathrm{W}(\gamma) =
	- \frac{1}{N} \, \int \diff \sigma  \tr \plexp 
	\brk*{ i \int _\s ^L A_m \, \diff x^m} 
	\brk*{ \widebar{\xi} \, \dx^m \Gamma_m \, \Psi }
	\plexp \brk*{ i \int _0 ^\s A_m \, \diff x^m} .
\end{align} 
If $\dx^m$ is light-like, the matrix coupling the supersymmetry parameter $\xi$ to the fermionic field
squares to zero 
\begin{align}
\left( \dx^m \Gamma_m \right) ^2 = \half \dx^m \dx^n \, 
	\left \lbrace \Gamma_m , \Gamma_n \right \rbrace = 0 \, ,
\end{align}
and thus its rank is at most half of its dimension. This implies that locally we can find at least 
sixteen linearly independent supersymmetry parameters $\xi (\sigma)$ for which the supersymmetry 
variation vanishes. A more careful analysis shows that the above restriction is compatible with the 
Majorana and Weyl conditions such that the Wilson loop is locally invariant under half of the 
supersymmetry transformations. 
Note, however, that the action is not invariant under local supersymmetry transformations 
such that our finding of local supersymmetry does not have immediate consequences in 
the form stated above. 

This property carries over to the counterpart of the light-like Wilson loop in $\mathcal{N}\!=4$ 
super Yang--Mills theory, the Maldacena--Wilson loop 
\begin{align}
W(\gamma) = \frac{1}{N} \, \tr \plexp 
	\brk*{ i \int _\gamma \brk*{ A_\mu \, \diff x^\mu 
		+ i \, \Phi_I \lvert\dx \rvert n^I } } .
\end{align}
Here, $n^I$ is a six-dimensional unit vector, which can in general depend on the curve parameter 
$\sigma$. This ensures that the constraint of light-like tangent vectors in ten dimensions is satisfied, 
\begin{align}
\dx ^m (\sigma) = \left( \dx^\mu(\sigma) , i \, n^I (\sigma) \lvert \dx (\sigma) \rvert \right) 
\quad \Rightarrow \quad
\dx ^m  \dx _m 
	= \dx^2 - \lvert \dx \rvert ^2 = 0 . 
\end{align}
Here, we have defined
\begin{align}
\lvert \dx \rvert = \begin{cases}
		\sqrt{\dx^2} \quad &\text{if} \; \; \dx^2 \geq 0 \, , \\
		i \, \sqrt{\lvert \dx^2 \rvert} \quad &\text{if} \; \; \dx^2 < 0 \, , 
	\end{cases}
\end{align}
such that the Maldacena--Wilson loop is only a phase if $\dx^\mu$ is time-like. For a space-like 
tangent vector, the additional components of the ten-dimensional vector are necessarily imaginary. 

The Maldacena--Wilson loop inherits the local supersymmetry property%
\footnote{We note that if $\dx^m$ has imaginary components, it is not possible to find solutions to 
$\dx^m \Gamma_m \, \xi = 0$, which satisfy the Majorana condition for spinors in ten dimensions. }
of the ten-dimensional Wilson loop. However, since the action is not invariant under local 
supersymmetry variations, only the invariance under global supersymmetry variations has implications
for the expectation value. The simplest case is the straight line for which the Maldacena--Wilson loop 
is a 1/2 BPS operator, such that its expectation value is finite and does not receive quantum
corrections,
\begin{align}
\left \langle W ( 
	\raisebox{1mm}{\includegraphics[width=2.2ex]{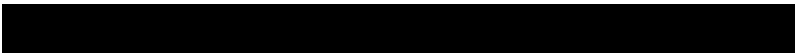}} 
	)  \right \rangle = 1 . 
\end{align}
We can understand the finiteness of the Maldacena--Wilson loop for smooth contours from this result. 
Recall that the divergences arise from the limit of all integration points being close to each other. 
In this limit, however, any curve behaves like a straight line since the curvature only gives a 
higher-order correction and the finiteness of the above expectation value thus carries over to 
generic smooth curves.  

If we take the sphere vector $n^I$ to be constant, only the straight line preserves some of the 
supersymmetry in the Euclidean case. In order to see this, consider the condition for supersymmetry, 
\begin{align}
(\Gamma^\mu \dx_\mu (\sigma) + i \Gamma^I n^I \lvert \dx \rvert  ) \xi = 0 . 
\label{eqn:susy-cond}
\end{align}
The loop preserves a fraction of the global supersymmetry, if there is a constant supersymmetry 
parameter $\xi$ satisfying the above condition for all points along the loop. Picking a parametrization 
for which $ \lvert \dx \rvert \equiv 1 $, we take the derivative of the above condition to find 
\begin{align*}
\Gamma^\mu \ddx_\mu (\sigma) \xi = 0 .
\end{align*}
This condition only has (local) solutions if $\ddx ^2 = 0$, since the matrix $\brk!{\Gamma^\mu \ddx_\mu}$ 
has non-vanishing determinant otherwise. In the Euclidean case, we are thus left with the straight line. 
For a Wilson loop in Minkowski space, there is also the option that $\ddx^\mu$ is light-like, which 
leads to a class of 1/4 BPS Maldacena--Wilson loops. The dual minimal surfaces are known, and indeed their area 
vanishes \cite{Ishizeki:2008dw}. 

Also in Euclidean space, however, there is a class of Maldacena--Wilson loops, which preserve some 
of the global supersymmetry. First, by introducing a coupling between the $\mathrm{S}^5$-vectors 
$n^I(\sigma)$ and the contour $x^\mu (\sigma)$, one may construct operators for which the supersymmetry 
condition
\begin{align*}
(\Gamma^\mu \dx_\mu (\sigma) + i \Gamma^I n^I (\sigma)) \xi = 0 
\end{align*}
does allow for constant solutions even if the above matrix is not constant \cite{Zarembo:2002an}.
Depending on the dimension of the subspace in which the curve can be embedded, different amounts 
of supersymmetry can be preserved leading to 1/4, 1/8 or 1/16 BPS operators. Moreover, one can also 
consider special superconformal symmetries in addition to the Poincar{\'e} supersymmetries discussed 
above. This leads to additional classes of contours \cite{Drukker:2007dw}. An important example of 
such a contour is the circular Wilson loop with constant sphere vector, for which the 
1/2 BPS symmetry was found in \cite{Bianchi:2002gz}. A classification of loops for which at least one 
supersymmetry can be preserved was obtained in \cite{Dymarsky:2009si,Cardinali:2012sy}. 

For explicitness, let us consider the expectation value of the Maldacena--Wilson loop at the one-loop 
order. Inserting the scalar propagator  
\begin{align*}
\left \langle \Phi _I ^a (x_1) \, \Phi _J ^b (x_2) \right \rangle 
	= \frac{g^2}{4 \pi^2} \,  
	\frac{\delta_{IJ} \, \delta ^{ab}}{(x_1-x_2)^2} \, ,
\end{align*}
we find 
\begin{align}
\left \langle W(\gamma) \right \rangle =
	1 - \frac{\lambda \left( 1 - N^{-2} \right)}{16 \pi ^2} \, 
	\int \diff \sigma_1 \, \diff \sigma_2 \, 
	\frac{\dx_1 \dx_2 - n_1 n_2 \, \lvert \dx_1 \rvert \lvert \dx_2 \rvert }
	{(x_1 - x_2)^2} + \Order (\lambda ^2) \, , 
\label{W1loop}	
\end{align}
Using this expression, it is easy to see that the one-loop result is indeed finite for a generic smooth curve.

For the Maldacena--Wilson loop, we have a generalized cusp anomalous dimension, depending on both
the angle $\phi$ of the cusp as well as the angle $\theta$ between the two $\mathrm{S}^5$ couplings before 
and after the cusp, $\cos \theta = n_1 \cdot n_2$. At the one-loop level, we can adapt the result for the 
cusp anomalous dimension of the Wilson loop using \eqref{W1loop} to find 
\begin{align}
\Gamma _{\text{cusp}} (\phi, \theta) \sim \frac{\phi \brk!{\cos \phi - \cos \theta} }{\sin \phi} . 
\end{align}
The vanishing of the cusp anomalous dimension in the case $\cos \phi = \cos \theta$ is not an 
accident and persists at all loop orders. This is an example of the class of 1/4 BPS Maldacena--Wilson 
loops that can be constructed in the plane following Zarembo's approach \cite{Zarembo:2002an}. The 
scalar coupling is related to the contour in such a way that locally around the cusp, the operator 
preserves some of the supersymmetry and hence the cusp anomaly is absent. 

Away from the BPS case, one encounters the Bremsstrahlung function $B(\lambda)$, 
which determines the energy emitted by a moving quark (hence the name) and appears 
in the small angle expansion as \cite{Correa:2012at}
\begin{align}
\Gamma _{\text{cusp}} (\phi, \theta) = \brk!{ \theta^2 - \phi^2 } B(\lambda) 
	+ \mathcal{O} \brk!{\theta^4} + \mathcal{O} \brk!{\phi^4} , 
\end{align}
and more generally in the expansion around the BPS configuration as
\begin{align}
\Gamma _{\text{cusp}} (\phi, \theta) = \brk!{ \theta - \phi } 
	\frac{2 \phi}{1- \ft{\phi^2}{\pi^2}} B\brk!{ \lambda \brk!{ 1 - \ft{\phi^2}{\pi^2} }} + \ldots  . 
\end{align}
The Bremsstrahlung function can be related to the expectation value of the Maldacena--Wilson 
loop over a circle \cite{Correa:2012at}. Since the latter can be calculated exactly (see section 
\ref{sec:circle}), also the Bremsstrahlung function is known as an exact function in both $\lambda$ 
and $N$. 

The cusp anomalous dimension can also be obtained from an integrability-based approach 
\cite{Drukker:2012de,Correa:2012hh,Drukker:2006xg}. In order to understand where integrability 
appears, it is helpful to map the two semi-infinte lines, which one typically considers for 
the cusp to a lens-shaped contour \cite{Drukker:2011za} containing an additional cusp of the same angle. In order to see this, one may e.g.\ consider the action of the inversion map 
$\mathrm{I}(x) ^\mu = \frac{x^\mu}{x^2}$ on two semi-infinte straight lines going out of 
the point $(0,1)$.

In this setup, we consider the insertion of the scalar fields $Z ^L$ and $\bar{Z}^L$ at the 
two opposite cusps. At a sufficient order in perturbation theory, we encounter e.g.\ the following 
diagram:
\begin{equation*}
\includegraphics{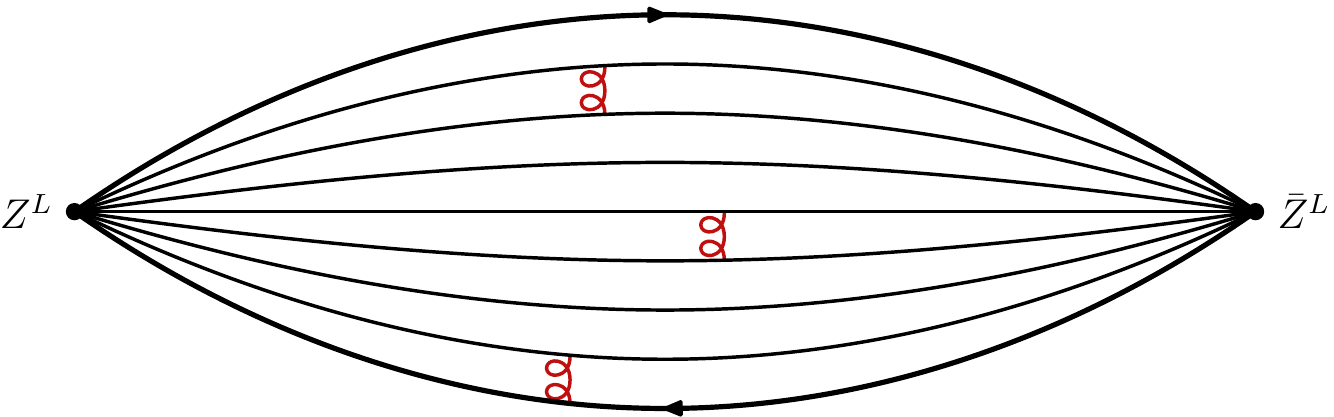}
\end{equation*}
Here, we may view the position of the scalar fields $Z$ as the sites of a spin chain. In this picture, 
the gluon propagators in the bulk of the diagram correspond to interactions between the sites (there 
are also other sources for interactions) and the lowest propagator corresponds to an interaction with 
the Wilson loop, which can be viewed as the boundary of the bulk spin chain.
The bulk spin chain is of course well known and the boundary reflection matrix following from the Wilson 
loop can be fixed by symmetry considerations \cite{Drukker:2012de,Correa:2012hh}. Finding the ground state 
energy in the limit $L \to 0$ then allows to extract the cusp anomalous dimension. Based on these ideas, 
modern techniques such as the Y-system and the quantum spectral curve have allowed to compute the cusp 
anomalous dimension with very high precision 
\cite{Gromov:2012eu,Gromov:2013qga,Gromov:2015dfa,Gromov:2016rrp}. 

In the case of a light-like cusp, which one obtains after analytically continuing $\phi \to i \gamma$ 
from the Euclidean cusp angle $\phi$ to a Minkowskian angle $\gamma$ and subsequently taking 
$\gamma \to \infty$, the anomalous dimension 
\begin{align}
\Gamma _{\text{cusp}} (\gamma,\lambda) = \gamma \Gamma _{\text{cusp}} (\lambda)
\end{align}
had been known before to allow for a integrability description known as the 
Beisert--Eden--Staudacher equation \cite{Eden:2006rx,Beisert:2006ez}. 

\subsection{Strong Coupling} 

On the string theory side of the AdS/CFT correspondence, the expectation value of the 
Maldacena--Wilson loop is given by the string partition function with the string configuration 
bounded by the Wilson loop contour on the conformal boundary of $\AdS_5$. In the limit of large 
coupling, the partition function is dominated by the smallest exponent, i.e.\ the minimal area 
that can be obtained given the boundary condition on the surface. The AdS/CFT prescription 
for the Maldacena--Wilson loop at strong coupling is hence given by \cite{Maldacena:1998im}
\begin{align}
\left \langle W(\gamma) \right \rangle \overset{\lambda \gg 1}{=}
	\exp \left( - \sfrac{\sqrt{\la}}{2 \pi} A_\mathrm{ren}(\gamma) \right) .
\label{W:Strong}	
\end{align}
Here, $A_\mathrm{ren}(\gamma)$ denotes the area of the minimal surface ending on the contour 
$\gamma$, which is situated at the conformal boundary. In order to describe the boundary value 
problem, we employ Poincar{\'e} coordinates $(X^\mu , y)$ for AdS, such that the metric is given by
\begin{align}
\diff s^2 = \frac{\diff X^\mu \, \diff X_\mu + \diff y \, \diff y}{y^2} \, . 
\end{align}
The conformal boundary corresponds to the surface at $y=0$. For suitably chosen coordinates $\tau$ 
and $\sigma$, we thus impose the boundary conditions
\begin{align}
X^\mu ( \tau = 0 , \sigma ) &= x^\mu (\sigma) \, , &
y ( \tau = 0 , \sigma ) &= 0 \, .
\end{align}
\begin{figure}
\begin{center}
\includegraphics{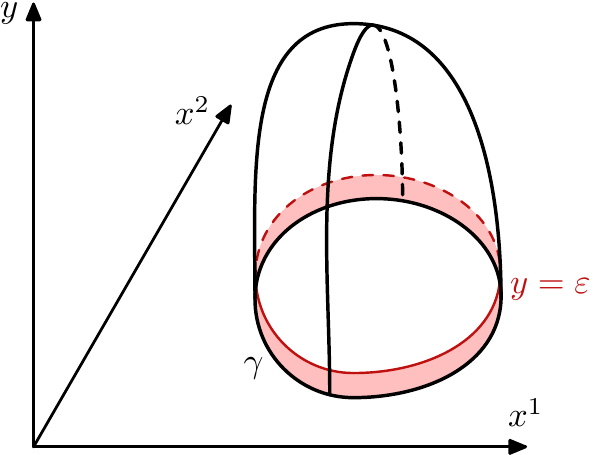}
\end{center}
\caption{Sketch of the minimal surface appearing in the strong-coupling description of the 
Maldacena--Wilson loop.}
\label{fig:Minimal-Surface}
\end{figure}
We can calculate the area of the minimal surface using either the Nambu--Goto or Polyakov action 
\begin{align}
A_{\text{NG}} & = \int \diff \tau \, \diff \sigma \, \sqrt{\det \left( \Gamma_{ab} \right) } \, , &
A_{\text{P}} & = \frac{1}{2} \int \diff \tau \, \diff \sigma \, \sqrt{h} h^{ab}  \Gamma_{ab}  \, ,
\end{align}
where $\Gamma_{ab} = y^{-2} \left( \partial_a X^\mu \, \partial_b X_\mu 
+ \partial_a y \, \partial_b y \right)$ is the induced metric on the surface. For the Polyakov action, 
we of course need to solve the equations of motion for the worldsheet metric $h$ first. 
In either case, however, there is a subtlety. Due to the divergence of the AdS-metric on the 
conformal boundary, the area of the minimal surface diverges as well. We can regulate it by introducing 
a cut-off $\varepsilon$	for the $y$-direction and integrating only over the region 
$y \geq \varepsilon$, see figure \ref{fig:Minimal-Surface}. 

Let us figure out how the area of the minimal surface diverges as we take $\varepsilon$ to zero. 
We expect the minimal surface to leave the boundary perpendicularly in order to avoid the regions 
where the metric is large. We can verify this expectation from the equations of motion directly. 
Note that this behaviour is unusual: Generically one would not expect to be able to derive an 
expansion around the boundary from the equations of motion, since they are underdetermined as an 
initial value problem. A unique solution only exists due to the second boundary condition; in our 
case this is the condition that the minimal surface closes.
In the case of a minimal surface ending on the conformal boundary of $\AdS_5$ however, the first 
few coefficients are fixed by the equations of motion and the undetermined ones are shifted to higher 
orders in the expansion around the boundary. 

Plugging a formal expansion into the equations of motion (in the Polyakov formalism and using 
conformal gauge), we find that \cite{Polyakov:2000jg,Polyakov:2000ti}
\begin{align}
X^\mu \left( \tau , \sigma \right) &= x^\mu (\sigma)
	+ \frac{\tau^2}{2} \, \dot{x}^2(\sigma) \, \partial_\sigma 
	\left( \frac{\dot{x}^\mu(\sigma)}{\dot{x}^2(\sigma)} \right) 
	+ \mathcal{O}\left(\tau^3 \right)  , \label{eqn:expansion_boundary} \\
y \left( \tau , \sigma \right) &= \tau \, \lvert \dot{x}(\sigma) \rvert 
	+ \mathcal{O}(\tau^3) .
\end{align}
This expansion is known as the Polyakov--Rychkov expansion. As expected, we see that the minimal 
surface leaves the boundary perpendicularly, since the first correction to $X^\mu (\sigma)$ 
appears only at the second order of the expansion. Taking into account the form of the metric, 
we thus see that the divergence is given by $L(\gamma)/\varepsilon$ (the reader is invited to 
confirm this by direct calculation) and note that the AdS/CFT prescription \eqref{W:Strong} contains 
the renormalized minimal area
\begin{align}
A_\mathrm{ren}(\gamma) = \lim \limits _{\varepsilon \to 0} 
	\left \lbrace A(\gamma) \big \vert_{y \geq \varepsilon} 
	- \frac{L(\gamma)}{\varepsilon} \right \rbrace .  
\label{def:Aren_0}
\end{align}
Note that the Maldacena--Wilson loop over a smooth contour is finite and does not
require renormalization --- at least not in addition to the field renormalization one would 
also have to consider in the calculation of amplitudes. The above renormalization of the area 
entails the contribution of the scalar fields at strong coupling in the case of constant $n^I$ 
which we have been considering so far. It stems from considering the Legendre transformation with 
respect to the loop variables coupling to the scalar fields, see reference \cite{Drukker:1999zq} 
for more details.
For the generic case of $n^I (\sigma)$ describing a closed curve on $\mathrm{S}^5$, the 
strong-coupling description contains a minimal surface in $\AdS_5 \times \mathrm{S}^5$, which is 
bounded by $x^\mu(\s)$ in the conformal boundary of $\AdS_5$ and $n^I(\sigma)$ in $\mathrm{S}^5$.
It is this additional piece that can lead to a vanishing of the total area for the BPS loops 
which have trivial expectation value. 

The third-order coefficient of $X^\mu$ is indeed not fixed by the equations of motion. We thus 
expect that it is related to the functional derivative of the minimal area. Let us thus consider 
the variation of the area given a variation $\delta x^\mu(\sigma)$ of the boundary curve. It
induces a variation $\left(\delta X^\mu , \delta y \right)$ of the parametrization of 
the minimal surface. The cut-off condition $y \geq \varepsilon$ translates to $\tau \geq \tau_0(\s)$
in parameter space, where $\tau_0(\s)$ is defined by $y(\tau_0(\s),\s) =\varepsilon$, which we 
can rewrite as 
\begin{align}
\tau_0(\sigma) = \frac{\varepsilon}{\lvert \dot{x} (\s) \rvert} + \Order ( \varepsilon^3) , 
\end{align}
employing the coefficients of $y$ derived above. Since we are varying around a minimal surface 
solution, we may employ that $(X^\mu,y)$ satisfy the equations of motion and hence the variation 
is given by a boundary term,
\begin{align}
\delta A \big \vert _{y \geq \varepsilon} 
	&= \int \limits _{y \geq \varepsilon} \diff \tau \, \diff \sigma 
	\, \partial_i \, \frac{ \partial_i X^\mu \, \delta X_\mu + \partial_i y \, \delta y}{y^2} 
	= \int \limits _0 ^{2 \pi} \diff \sigma \hspace*{-2mm} 
	\int \limits _{\tau_0(\sigma)} ^c  \hspace*{-2mm} \diff \tau 
	\, \partial_i \, \frac{ \partial_i X^\mu \, \delta X_\mu + \partial_i y \, \delta y}{y^2} \\
	&= \frac{1}{\varepsilon^2} \int \limits _0 ^{2 \pi} \diff \sigma 
	\left[ \tau_0^\prime (\sigma) \,  \partial_\sigma X^\mu \delta X_\mu 
	-  \partial_\tau X^\mu \delta X_\mu \right] .
\end{align}
Here, we used that $\delta y(\tau_0(\s),\s)=0$ due to the definition of $\tau_0$ and employed 
the periodicity of the solutions in $\sigma$. Inserting the results \eqref{eqn:expansion_boundary}, 
we then find
\begin{align*}
\delta A \big\vert _{y \geq \varepsilon} = \frac{\delta L(\gamma)}{\varepsilon} 
	- \int \limits _0 ^{2 \pi}
	\diff \sigma \, \frac{3 X_{(3)}^\mu}{\dx^2} \, \delta x_\mu \, ,
\end{align*}
from which we read off that
\begin{align}
X_{(3)} ^\mu (\s) = - \frac{\dx^2}{3} \, \dfrac{\delta A_\mathrm{ren}(\gamma) }
	{\delta x_\mu (\sigma) } . \label{funcder}
\end{align}
The third-order coefficient of $y$ can be determined from the Virasoro constraints. The expansion
then reads 
\begin{align}
X^\mu \left( \tau , \sigma \right) &= x^\mu (\sigma)
	+ \frac{\tau^2}{2} \, \ddot{x}^\mu (\sigma)
	- \frac{\tau^3}{3} \, \frac{\delta A_{\mathrm{ren}}(\gamma)}{\delta x_\mu (\sigma)} 
	+ \Order \left(\tau^4 \right)  ,\label{eqn:expansion_boundary_arc} \\
y \left( \tau , \sigma \right) &= \tau \,  
	- \frac{\tau^3}{3} \, \ddot{x}(\sigma) ^2 + \Order (\tau^4) .
\end{align}
Here, we have fixed the parametrization of the boundary curve to satisfy $\dx ^2 = 1$ in order 
to simplify the expansion. 

We may also employ the above expansion in order to show that the area of the minimal surface 
is invariant under conformal transformations following an argument given in reference 
\cite{Dorn:2015bfa}. Since the conformal transformations are the boundary limits of isometries 
of $\AdS_5$, it is clear that the transformation of the minimal surface associated to the 
transformation of the boundary curve is a symmetry of the area functional. We should, however, 
also consider that the non-renormalized area is divergent. Indeed, the coefficient of this divergence is 
not invariant under conformal transformations. 

The point here is that the transformation of the surface cut off at $y = \varepsilon$ does 
not lead to another surface that is cut-off in the same way. The difference between the 
original surface $(X^\mu(\tau,\sigma),y(\tau,\sigma))$ 
and the transformed surface $(\tilde{X}^\mu(\tau,\sigma),\tilde{y}(\tau,\sigma))$ 
thus arises from the integration over the region 
between the two cut-offs situated at $\tau_0 (\sigma)$ and $\tilde{\tau}_0 (\sigma)$. 
Again employing the Polyakov--Rychkov expansion, we find this difference to be given by  
\begin{align}
& A_{\mathrm{min}}(\gamma) \big \vert _{y \geq \varepsilon} 
	- A_{\mathrm{min}}(\tilde{\gamma}) \big \vert _{\tilde{y} \geq \varepsilon} 
	=  \frac{1}{2} \int \limits _{0} ^{2 \pi} \diff \sigma 
	\int \limits _{\tau_0 (\sigma)} ^{\tilde{\tau}_0 (\sigma)} \diff \tau \, 
	\frac{\partial_i \tilde{X}^\mu \partial_i \tilde{X}^\mu 
	+\partial_i \tilde{y} \partial_i \tilde{y}}{\tilde{y}^2} \nn \\
	& \qquad = \int \limits _{0} ^{2 \pi} \diff \sigma 
	\int \limits _{\tau_0 (\sigma)} ^{\tilde{\tau}_0 (\sigma)} \diff \tau \, 
	\left( \frac{1}{\tau^2} + \mathcal{O} ({\tau^0}) \right) \nn \\
	& \qquad =   \int \limits _{0} ^{2 \pi} \diff \sigma \, 
	\frac{\lvert \dot{x}(\sigma )\rvert }{\varepsilon} 
	- \int \limits _{0} ^{2 \pi} \diff \sigma \, 
	\frac{ \lvert \dot{\tilde{x}}(\sigma ) \rvert}{\varepsilon} + \mathcal{O}(\varepsilon) 
	= \frac{L(\gamma)}{\varepsilon} - \frac{L(\tilde{\gamma})}{\varepsilon} 
	+ \mathcal{O}(\varepsilon) .
\end{align}
This shows that the renormalized area \eqref{def:Aren_0} is indeed invariant. Note that the argument 
given here applies to any symmetry of the area functional, they need not be isometries of $\AdS_5$. 

\subsection{Circular Maldacena--Wilson loop}
\label{sec:circle}
One contour of particular interest within the AdS/CFT correspondence is the circle, for which 
the expectation value of the Maldacena--Wilson loop can be calculated exactly on the gauge theory side, 
thus allowing for a comparison with the AdS/CFT prediction at strong coupling.
The circular Maldacena--Wilson loop is a 1/2 BPS operator, if one considers also the superconformal 
symmetries of the theory \cite{Drukker:2007dw}. Incidentally, the expectation value is not trivial 
since the variations are not pure supersymmetries.  

The minimal surface for the circular Wilson loop in Euclidean space was obtained soon after the AdS/CFT 
proposal in reference \cite{Berenstein:1998ij}. It is natural to assume that the sections of the minimal 
surface at constant $y$ are still circular. Hence, we consider the ansatz
\begin{align}
X^\mu (r , \sigma ) &= \left( r \cos \sigma , r \sin \sigma \right) \, , &
y &= y(r) . 
\label{SurfCircleInitial}
\end{align} 
The Nambu--Goto action then gives the area
\begin{align}
A = \int \diff r \, \diff \sigma \, \frac{r \sqrt{1+ y^\prime (r) ^2}}{y(r)^2} \, , 
\end{align}
such that we have the equations of motion
\begin{align}
\partial_r \left( \frac{r \, y^\prime (r)}{y(r) ^2 \sqrt{1+ y^\prime (r) ^2} } \right) 
	+ \frac{2r \, \sqrt{1+ y^\prime (r) ^2}}{y(r)^3} = 0 , 
\label{Circle:EOM}	
\end{align}
along with the boundary condition $y(1)=0$ for a circle of radius 1. Even though the problem of finding 
the minimal surface has simplified to an ordinary differential equation, it is still non-trivial to solve. 
We may, however, obtain the solution by using that the inversion map on $\mathbb{R}^2$,
\begin{align*}
I(x)^\mu = \frac{x^\mu}{x^2} \, ,
\end{align*}
maps the circle to a straight line and vice versa. To be precise, consider the curves
\begin{align*}
x(\sigma) &= \left( \cos \sigma , \sin \sigma + 1 \right) \, , &
I(x(\sigma)) &= \left( \frac{\cos \sigma}{2 ( 1 + \sin \sigma)}  , \frac{1}{2} \right) .
\end{align*} 
The inversion map can be extended to the AdS-isometry 
\begin{align}
I_\AdS (X,y) &= \left( \frac{X^\mu}{X^2 + y^2} \, , \, \frac{y}{X^2 + y^2} \right) ,
\end{align}
which can be used to map the (formal) minimal surface attached to the straight line to the one attached to the 
circle. We may write the minimal surface for the straight-line as
\begin{align}
X^\mu (\tau , \sigma) &= \left( \sigma , \ft{1}{2} \right) \, , &
y(\tau , \sigma ) &= \tau \, . 
\label{Line-Surface}  
\end{align}
It is a straightforward exercise to check that this surface gives a solution of the equations 
of motion. After employing the inversion map in $\AdS_5$, we obtain the surface
\begin{align}
X^\mu (\tau , \sigma ) &= \left( \frac{4 \sigma}{1 + 4 ( \sigma ^2 + \tau ^2 )} \, , 
	\frac{1 - 4 (\sigma ^2 + \tau ^2) }{1 + 4 ( \sigma ^2 + \tau ^2 ) } \right)  \, , &
y (\tau , \sigma) = \frac{4 \tau}{1 + 4 ( \sigma ^2 + \tau ^2 )} .	
\label{Surf:Circle1}
\end{align}
Here, we have employed a translation by $(0,-1)$ in addition to the inversion such that the circle is centered 
around the origin. Even though it satisfies conformal gauge (since \eqref{Line-Surface} does), the 
parametrization obtained above is not particularly simple. In order to reach the form of our ansatz 
\eqref{SurfCircleInitial}, note that the surface described by equation \eqref{Surf:Circle1} satisfies the equation
\begin{align}
X^2 + y^2 = 1 . 
\end{align}
For our original parametrization \eqref{SurfCircleInitial}, we thus find
\begin{align}
y(r) = \sqrt{1 - r^2} \, ,
\end{align}
which indeed solves the equations of motion \eqref{Circle:EOM}. Another often-used parametrization is given by
\begin{align}
X_1 (\tau , \sigma) &= \frac{\cos \sigma}{\cosh \tau} \, , &
X_2 (\tau , \sigma) &= \frac{\sin \sigma}{\cosh \tau} \, , & 
y (\tau , \sigma) &= \tanh \tau . &  
\label{sol:circle:0} 
\end{align} 
For this parametrization, the induced metric is Weyl-equivalent to the flat metric as well, such that 
it solves the equations of motion following from the Polyakov action in conformal gauge.

In order to calculate the area of the minimal surface, we introduce a cut-off at 
$y= \varepsilon$, corresponding to $r= \sqrt{1 - \varepsilon^2}$, and obtain
\begin{align}
A_{\mathrm{ren}} \left( 
	\raisebox{-.9mm}{\includegraphics[height=2.2ex]{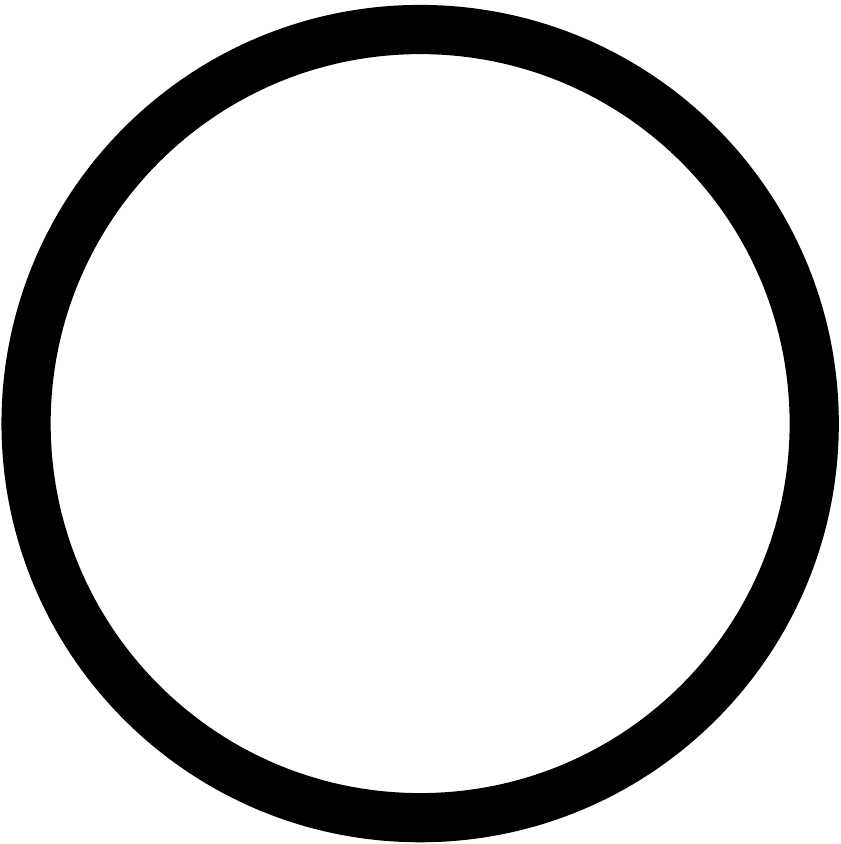}}   
	\right)
	= \lim \limits _{\varepsilon \to 0} 
	\bigg \lbrace \int \limits _0 ^{2 \pi} \diff \sigma 
	\int \limits _0 ^{\sqrt{1-\varepsilon^2}} 
	\frac{r \, \diff r}{\left( 1 - r^2 \right)^{3/2} } \, 
	- \frac{2 \pi}{\varepsilon}  \bigg \rbrace
	= - 2 \pi .
\end{align} 
We have thus found that the circular Maldacena--Wilson loop has the following asymptotic behavior at strong coupling:
\begin{align}
\left \langle W  \left( 
	\raisebox{-.9mm}{\includegraphics[height=2.2ex]{circle.pdf}}   
	\right) 
	\right \rangle 
	\overset{\lambda \gg 1}{=}  
	e^{\sqrt{\lambda}} .
\label{eqn:Pred}	
\end{align}

We note that the area of the minimal surfaces is always negative, 
i.e.\ the finite correction to the leading term $L(\gamma)/\varepsilon$ is negative. It is an
interesting exercise to show that this is the case for any smooth contour. 

\begin{figure}[t]
\centering
\subcaptionbox{Double-Propagator}
[.3\linewidth]{\includegraphics[width=.2\linewidth]{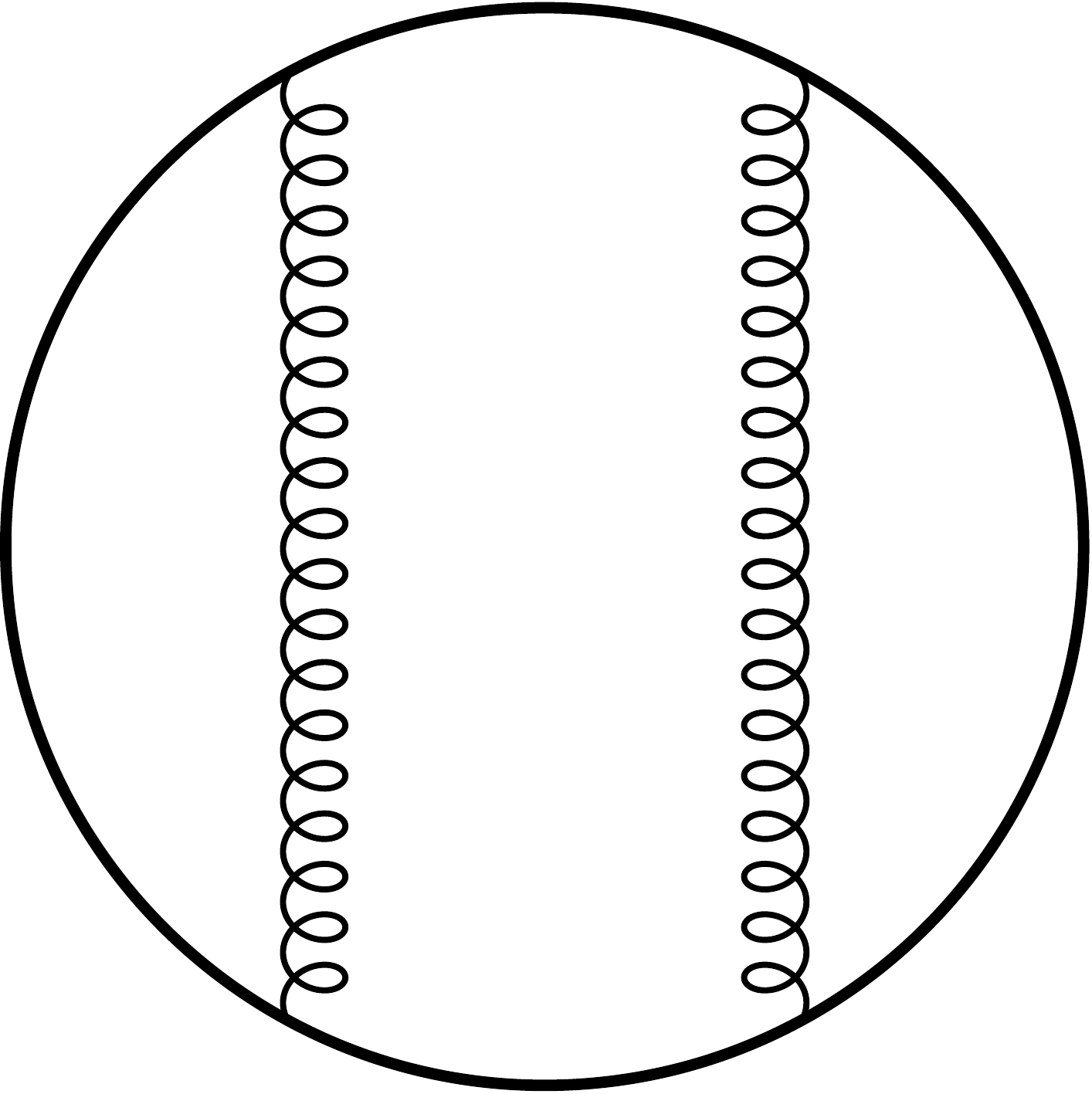}}
\subcaptionbox{Self-Energy} 
[.3\linewidth]{\includegraphics[width=.2\linewidth]{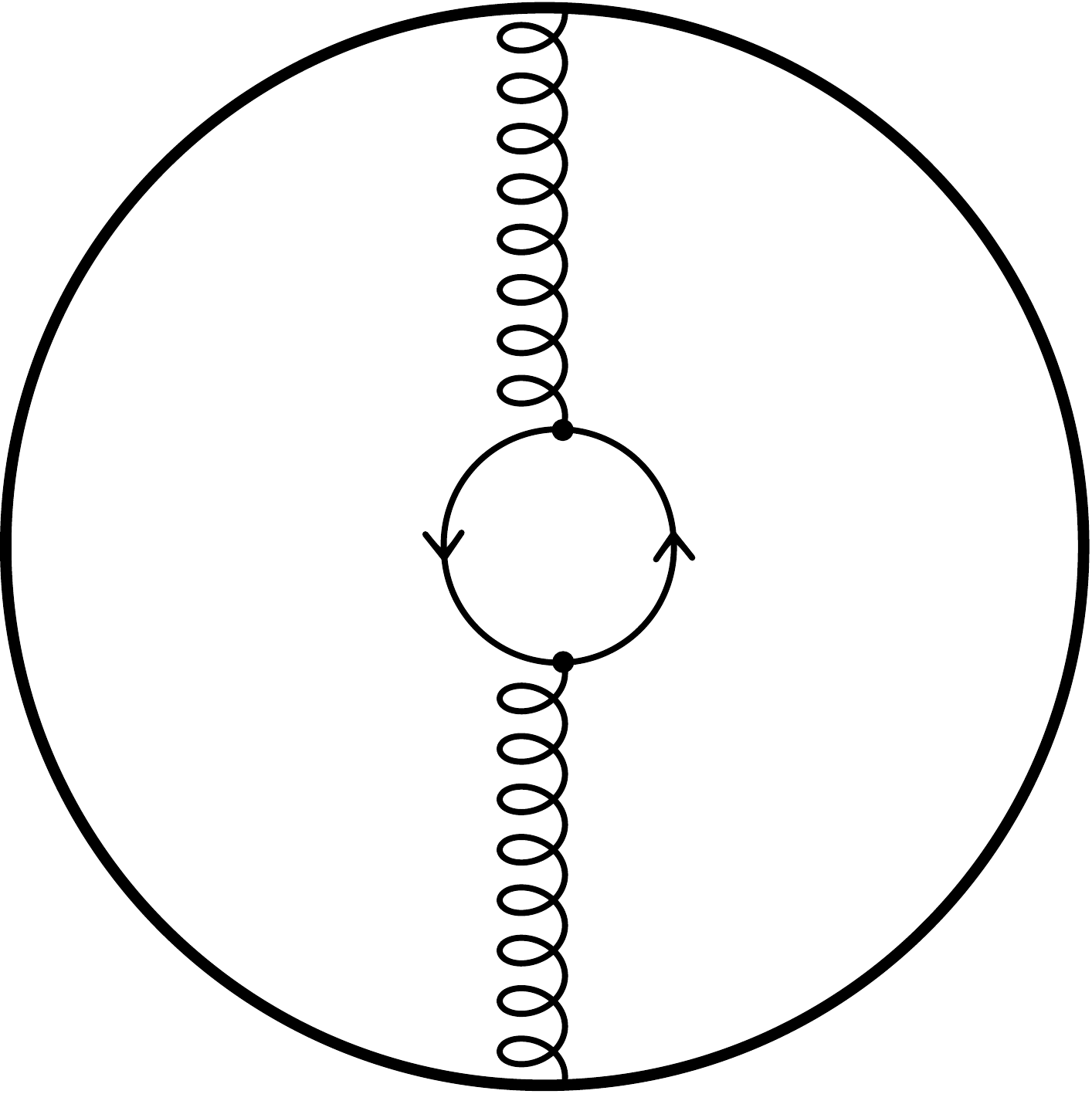}}
\subcaptionbox{Three-Vertex} 
[.3\linewidth]{\includegraphics[width=.2\linewidth]{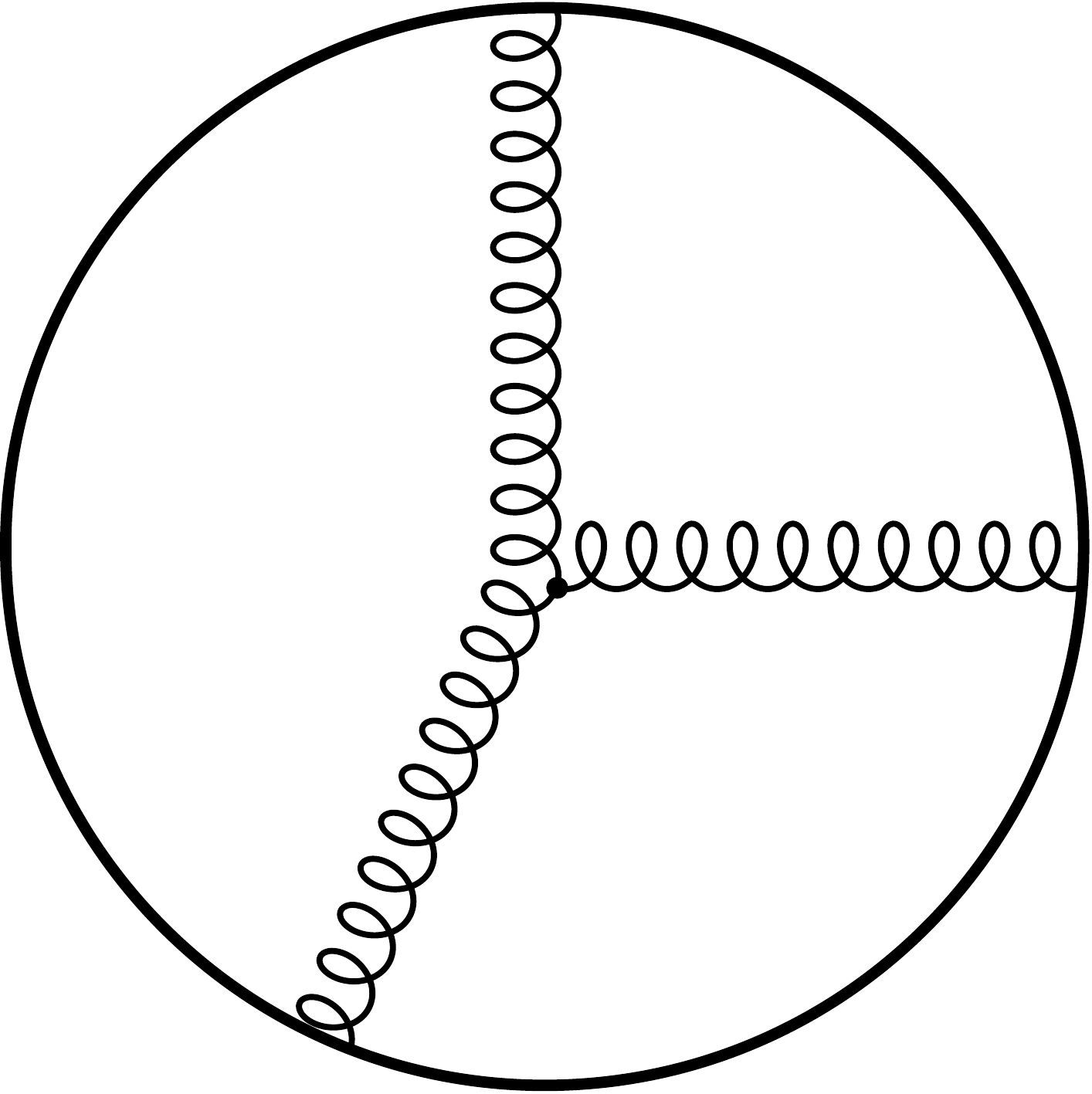}}
\caption{Examples of the double-propagator, self-energy and three-vertex diagrams appearing in the two-loop calculation of the expectation value of the Maldacena--Wilson loop. }
\label{fig:2-loop-diagrams}
\end{figure}

It is a remarkable achievement that the expectation value of the circular Maldacena--Wilson loop has 
been calculated exactly on the gauge theory side, beginning with the calculation of reference 
\cite{Erickson:2000af}, which is sketched below. Let us first consider the one-loop order of the expectation 
value \eqref{W1loop}. For the circle parametrized by $x(\sigma) = ( \cos \sigma , \sin \sigma)$ and $n^I$ 
constant, we find
\begin{align}
\frac{\dx_1 \dx _2 - \vert \dx_1 \vert \vert \dx_2 \vert}{(x_1 - x_2 ) ^2}
	= \frac{\cos \sigma_1 \cos \sigma _2  + \sin \sigma_1 \sin \sigma_2 - 1 }
	{2 - 2 ( \cos \sigma_1 \cos \sigma _2  + \sin \sigma_1 \sin \sigma_2) }
	= - \frac{1}{2} .
\label{prop-triv}	
\end{align}
The integral is hence trivial and in the planar limit, we obtain
\begin{align}
\left \langle W  \left( 
	\raisebox{-.9mm}{\includegraphics[height=2.2ex]{circle.pdf}}   
	\right) \right \rangle
	= 1 + \frac{\lambda}{8} + \Order( \lambda ^2 ) .	
\end{align}
At the next loop order, we need to take diagrams with three-vertices and the self-energy correction into 
account as well. The different types of diagrams are shown in figure \ref{fig:2-loop-diagrams}. 
These diagrams are divergent and require regularization. 
In a supersymmetric theory, it is convenient to employ the dimensional reduction scheme \cite{Siegel:1979wq}
as in the original calculation in reference \cite{Erickson:2000af}. This scheme is a version of dimensional 
regularization, in which $\mathcal{N} \! =  4$ supersymmetric Yang--Mills theory is viewed as the theory 
obtained from dimensionally reducing ten-dimensional $\mathcal{N}=1$ supersymmetric Yang--Mills theory to 
D dimensions. The regularized theory hence has a D-component vector field $A_\mu ^a$ as well as $10 - \mathrm{D}$ 
scalar fields $\Phi _I ^a$. Note that the expectation value of the Maldacena--Wilson loop remains finite even though 
some of the contributing diagrams diverge individually. Indeed, one observes that the divergences of the 
self-energy and three-vertex diagrams cancel each other for generic (smooth) contours. In the case of the circle, 
this cancellation is exact and hence the two-loop result comes only from propagators along the loop, which again 
lead to trivial integrals as in \eqref{prop-triv}. 

The calculation of reference \cite{Erickson:2000af} is now based on the conjecture that similar cancellations occur 
at all loop orders, such that the result can be calculated from diagrams without internal vertices. 
Given this conjecture, we only consider diagrams containing propagators ending on the circle, such as diagram (a) 
in figure \ref{fig:2-loop-diagrams}. Moreover, for the leading contributions in the planar limit, the propagators 
do not cross each other. These diagrams can easily be calculated, if we again combine the gluon and scalar 
contributions as in the one-loop calculation. Then each propagator contributes a factor of 
\begin{align*}
- \frac{\dx_1 \dx _2 - \vert \dx_1 \vert \vert \dx_2 \vert}{(x_1 - x_2 ) ^2} = \frac{1}{2} .
\end{align*}
The color factors follow from repeatedly employing the identity $t^a t^a = \ft{N}{2} \unit$ and along 
with the result 
\begin{align*}
\frac{(2 \pi)^{2n}}{(2n)!}
\end{align*}
for the ordered 2n-fold integral over the interval $[ 0 , 2 \pi]$, we find the contribution
\begin{align}
\frac{1}{2^n} \, \frac{(2 \pi)^{2n}}{(2n)!} \, \left( \frac{g^2}{4 \pi ^2} \right)^n
	\left( \frac{N}{2} \right) ^n = \frac{\lambda^n}{4 ^n \, (2n)!} 
\label{eqn:Diagramfactor}	
\end{align}
for each individual diagram at the $n$-th loop order. We need thus only count all possible rainbow-like diagrams 
consisting of $n$ propagators which are not crossing each other.
In order to find this number, note that any rainbow-like diagram with $n$ propagators contains 
such diagrams with less propagators, e.g.\ we might have the form
\begin{align*}
\includegraphics[width=82mm]{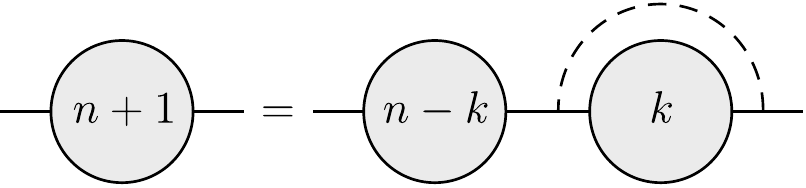} \, ,
\end{align*}
where the grey blob denotes a generic rainbow-like diagram containing the number of propagators
indicated. It is then easy to see that the number $A_n$ of the rainbow-like diagrams satisfies the 
recursion relation
\begin{align}
A_{n+1} &= \sum \limits _{k=0} ^n A_{n-k} \, A_k \, , &
A_0 &= 1 .
\label{eqn:recursion}
\end{align}
For the generating function  
\begin{align*}
f(z) = \sum \limits _{n=0} ^\infty A_n \, z^n \, ,
\end{align*}
the recursion relation turns into the functional equation 
\begin{align}
f(z) ^2 &= \frac{f(z) - 1 }{z} \, , &
f(0) &= 1 \, ,
\end{align}
which is solved by
\begin{align}
f(z) = \frac{1 - \sqrt{1-4z}}{2z} = \sum \limits _{n=0} ^\infty
	\frac{(2n)!}{(n+1)! \, n!} \, z^n .
\end{align} 
The number of rainbow-like diagrams containing $n$ propagators is hence given by
\begin{align}
A_n = \frac{(2n)!}{(n+1)! \, n!}  .
\end{align}
It was noted in reference \cite{Erickson:2000af} that $A_n$ can also be calculated from a matrix model introduced 
in reference \cite{Brezin:1977sv}. Combining the above finding with the factor \eqref{eqn:Diagramfactor} 
contributed by each individual diagram gives
\begin{align}
\left \langle W  \left( 
	\raisebox{-.9mm}{\includegraphics[height=2.2ex]{circle.pdf}}   
	\right) \right \rangle
	= \sum \limits _{n=0} ^\infty 
	\frac{\lambda ^n}{4^n \, (n+1)! \, n!} 
	= \frac{2}{\sqrt{\lambda}} \, \mathrm{I}_1 \big( \sqrt{\lambda} \, \big) .
\end{align}
Here, $\mathrm{I}_1$ is a modified or hyperbolic Bessel function of the first kind, cf.\ e.g.\ 
reference \cite{FunctionAtlas} for more details. The asymptotic expansion for large $\lambda$ is given by
\begin{align}
\left \langle W  \left( 
	\raisebox{-.9mm}{\includegraphics[height=2.2ex]{circle.pdf}}   
	\right) \right \rangle
	\overset{\la \gg 1}{=} \sqrt{\frac{2}{\pi}} \, 
	\frac{e^{\sqrt{\lambda}}}{\lambda^{3/4}} .  
\end{align}
It agrees with the AdS/CFT prediction \eqref{eqn:Pred} within the limits of its accuracy. 

The above calculation was extended by Drukker and Gross \cite{Drukker:2000rr} to include all 
non-planar corrections by studying the anomaly arising from the singular mapping of the straight 
line to the circle. They also relied on the conjecture that all diagrams containing interaction 
vertices cancel against each other. This conjecture was later proven by Pestun \cite{Pestun:2007rz}, 
who used localization techniques to reduce the calculation of the circular Maldacena--Wilson loop 
to a matrix model calculation, cf. also the reviews \cite{Pestun:2016zxk,Zarembo:2016bbk}.

At the strong-coupling side, extending the result beyond the classical area of the minimal surface 
proved difficult due to several ambiguities in the formalism for calculating one-loop correction to 
the partition function. The mismatches observed in the first of these calculations 
\cite{Drukker:2000ep,Kruczenski:2008zk} were attributed to these ambiguities, cf.\ e.g.\ reference 
\cite{Vescovi:2016zzu} for more details. In this light, it is interesting to consider the ratio between 
the circular 1/2 BPS Maldacena--Wilson loop and a 1/4 BPS Maldacena--Wilson loop known as the latitude 
Wilson loop, for which some of the potential ambiguities of the string one-loop calculation drop out.
The mismatch between the localization result and the string correction observed 
there \cite{Forini:2015bgo,Faraggi:2016ekd} could recently be resolved
\cite{Forini:2017whz,Cagnazzo:2017sny,Medina-Rincon:2018wjs}.

\subsection{Duality to Scattering Amplitudes}

In the discussion of the UV divergences of Wilson loops, we have noted that the anomalous dimension 
matrix for a self-intersecting Wilson loop happens to describe the IR divergences of a related 
scattering amplitude. In the simplest case, the IR divergences are described entirely by the cusp 
anomalous dimension. In a planar theory, this behavior extends to many-particle scattering amplitudes, 
if we consider color-ordered (partial) amplitudes, see references \cite{Dixon:1996wi,Henn:2014yza} 
for an introduction. For these, IR divergences exclusively stem from adjacent particles and are described 
by the cusp anomalous dimension. 

\begin{figure}[th]
\centering
\includegraphics{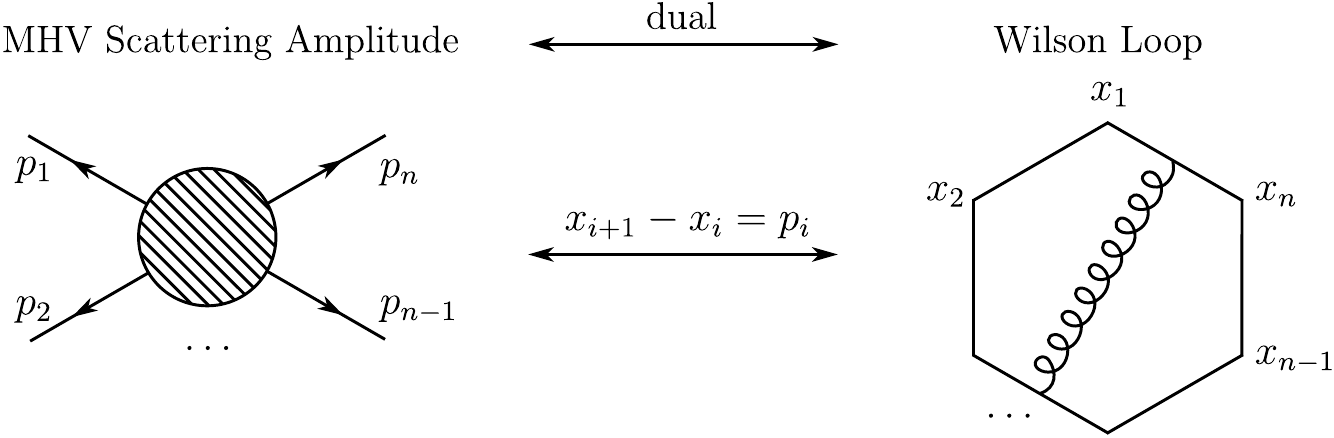}
\caption{Graphical representation of the duality between Wilson loops and scattering amplitudes.}
\label{fig:Duality}
\end{figure}

In $\mathcal{N}\!=4$ supersymmetric Yang--Mills theory, the connection between scattering amplitudes and 
Wilson loops goes even further, cf.\ the reviews \cite{Alday:2008yw,Henn:2009bd} for a more detailed 
discussion of the ideas sketched below. The first signs of the conjectured duality were observed by Alday 
and Maldacena \cite{Alday:2007hr}, who found that maximally helicity violating (MHV) gluon scattering 
amplitudes at strong coupling are described by the area of certain minimal surfaces and hence identical 
to the Maldacena--Wilson loop over the respective boundary contour. Concretely, the boundary curves are 
given by polygons with light-like edges, with the following relation between the cusp points $x_i$ and gluon 
momenta $p_i$: 
\begin{align}
x_{i+1} - x_i = p_i .  
\end{align} 
The leading behavior of these amplitudes at strong coupling is hence described by the Wilson loop over 
the polygon with the above cusp points, see also figure \ref{fig:Duality}. 

Let us shortly explain the nature of the duality in more detail. For the MHV amplitudes we are considering, 
two of the gluons have one helicity while all other gluons have the opposite helicity. In this case, the 
same function of the helicity variables appears at all loop orders, and the amplitude can be written as
\begin{align}
A_n = A_n ^{\mathrm{tree}} \, M_n \, . 
\end{align}
Here, $M_n$ is a function only of the momentum invariants $(p_i+p_j)^2$, the information on the helicity of 
the particles in entirely contained in the tree-level amplitude $A_n ^{\mathrm{tree}}$. The conjectured duality 
states \cite{Brandhuber:2007yx} that the function $M_n$ is equal%
\footnote{Since the duality relates ultraviolet and infrared divergent quantities, both the regularization parameters $\epsilon_{\mathrm{UV}}$ and $\epsilon_{\mathrm{IR}}$ and the renormalization constants $\mu_{\mathrm{UV}}$ and $\mu_{\mathrm{IR}}$ have to be related to each other. This can in general be done in such a way that the divergent pieces of the amplitude and the Wilson loop match, cf.\ e.g.\ reference \cite{Henn:2009bd} for a more detailed explanation.}
to the expectation value $\left \langle W_n \right \rangle$ of the related Wilson loop up to a constant $d$, 
\begin{align}
\left \langle W_n \right \rangle = M_n + d .
\end{align}

The duality is also of interest since the polygonal Wilson loops can be approached from an integrability 
calculation via a thermodynamic Bethe ansatz \cite{Alday:2010vh}. 
An important check of the duality is the case of six particles or cusps, respectively. This number of 
momenta is important for the following reason: From the viewpoint of the scattering amplitudes, the conformal 
symmetry of the Wilson loop appears as a dual conformal symmetry in the dual variables $x_i$; this symmetry 
had indeed been observed in calculations of scattering amplitudes \cite{Drummond:2006rz,Drummond:2007aua}.
The dual conformal symmetry is very restrictive in the case of four- and five-point scattering amplitudes, 
since there is no way to construct conformally invariant combinations of the dual variables $x_i$, due to 
the constraint that the points be light-like separated. This changes starting from six points, for which we can 
e.g.\ construct three conformally invariant cross-ratios
\begin{align*}
u_1 &= \frac{x_{13}^2 x_{46}^2}{x_{14}^2 x_{36}^2} \, , & 
u_2 &= \frac{x_{24}^2 x_{15}^2}{x_{25}^2 x_{14}^2} \, , &
u_3 &= \frac{x_{35}^2 x_{26}^2}{x_{36}^2 x_{25}^2} \, .
\end{align*}
In addition to passing checks for a lesser number of momenta \cite{Drummond:2007au,Drummond:2007cf}, the 
conjecture was also found to hold true for six points and two loops 
\cite{Drummond:2007bm,Bern:2008ap,Drummond:2008aq}. Here, both the Wilson loop and the scattering amplitude 
begin to deviate from an earlier conjecture known as the BDS ansatz \cite{Bern:2005iz} and start to depend 
on the above cross-ratios in a non-trivial way.  

\section{Integrability and Minimal Surfaces}
\label{sec:Symm}

The most immediate way in which integrability appears in the correspondence between 
$\mathcal{N}\! = 4 $ super Yang--Mills theory and string theory in $\AdS_5 \times \mathrm{S}^5$
is via the classical integrability of the string theory. This setup applies to the minimal surfaces 
appearing in the strong-coupling description of the Maldacena--Wilson loop and they are thus 
a natural starting point for exploring integrability in the context of the Maldacena--Wilson loops.  
Below, we will exploit the integrability of the minimal surface problem in order to derive hidden
symmetries for the Maldacena--Wilson loop at strong coupling. 

\subsection{Symmetric Space Models}
Before we turn to the discussion of minimal surfaces in $\AdS_5$, we briefly introduce a group-theoretic 
formalism to efficiently work with string models on symmetric spaces. More detailed introductions 
can be found in references \cite{Zarembo:2017muf,Munkler:2017rft}. 

Recall first that a homogeneous space $\grp{M}$ can be identified with the coset space obtained by dividing 
the isotropy group $\grp{H}$ of a point out of the isometry group $\grp{G}$ of the space, 
\begin{align}
\grp{M} \simeq \frac{\grp{G}}{\grp{H}} \, . 
\end{align}
For the case of a symmetric space, the Lie algebra of the isometry group $\grp{G}$ can additionally be 
decomposed as
\begin{align}
\alg{g} = \alg{h} \oplus \alg{m} 
\end{align}
in such a way that the constituents satisfy the relations (also known as a $\mathbb{Z}_2$-grading)   
\begin{align}
\brk[s]*{\alg{h} , \alg{h}} &\subset \alg{h} \, ,&
\brk[s]*{\alg{h} , \alg{m}} &\subset \alg{m} \, ,&
\brk[s]*{\alg{m} , \alg{m}} &\subset \alg{h} \, .
\end{align}
While the first two relations are related to $\alg{h}$ forming a Lie subalgebra, the latter relation 
does not follow generically and is only valid for symmetric spaces. 

The formalism is based on the Maurer--Cartan form
\begin{align*}
U = g^{-1} \diff g = A + a , 
\end{align*}
which takes values in the Lie algebra $\alg{g}$. Here, $A$ and $a$ denote the projections of the 
Maurer--Cartan form $U$ on $\alg{h}$ and $\alg{m}$, respectively. The metric for the coset space is then obtained 
from the group metric (which we denote by the trace here) applied to the projection $a$ of the Maurer--Cartan form, 
\begin{align}
\Gamma_{ij} = \tr \left( a _i \, a_j \right) . 
\end{align}
We can thus write the Polyakov action as
\begin{align}
A_P = \frac{1}{2} \int \diff ^2 \sigma \sqrt{h} h^{ij} \tr \left( a _i \, a_j \right) .
\end{align}
In the following, we will assume a Euclidean signature of the worldsheet metric, which is 
appropriate for the minimal surfaces we consider. The above group-theoretic formalism is particularly 
well-suited for the study of symmetries using a Lax connection. But first, let us see how to represent 
$\AdS_5$ in the way discussed above. 

\subsection{The Coset Construction for AdS}
In the case of $\AdS_5$\footnote{We consider $\AdS_5$ for explicitness. However, the construction given here applies to any 
dimension and also to Euclidean signature. }, we use the form
\begin{align*}
\AdS_5 \simeq \frac{\grp{SO}(2,4)}{\grp{SO}(1,4)} .
\end{align*}
For the generators $\lbrace P_\mu , M_{\mu \nu} , D , K_\mu \rbrace$ of the isometry group 
$\grp{SO}(2,4)$, we note the commutation relations 
\begin{align}
\left[M_{\mu \nu} , M_{\rho \sigma} \right] = \eta_{\mu \rho} M_{\nu \sigma} 
	- \eta_{\mu \sigma} M_{\nu \rho} + \eta_{\nu \sigma} M_{\mu \rho} 
	- \eta_{\nu \rho} M_{\mu \sigma} \, , 
\end{align}
as well as 
\begin{align}
\left[ D, P_{\mu} \right] &=  P_{\mu} \, ,&  
\left[ M_{\mu \nu}, P_{\lambda} \right] &= \eta_{\mu \lambda} P_{\nu}  
	- \eta_{\nu \lambda} P_{\mu}   \, , & 
\left[ P_{\mu}, K_{\nu} \right] &=  2 \eta_{\mu \nu} \, D  - 2 M_{\mu \nu} \, , \nn \\
\left[ D, K_{\mu} \right] &= - K_{\mu}\, ,&  
\left[ M_{\mu \nu}, K_{\lambda} \right] &= \eta_{\mu \lambda} K_{\nu} 
	-  \eta_{\nu \lambda} K_{\mu} .   \label{conf_algebra}
\end{align}
Moreover, we will also need the trace metric for which we note 
\begin{align}
\tr \left( M_{\mu \nu} M_{\rho \sigma} \right) &= 
	2 \, \eta_{\mu \sigma} \, \eta_{\nu \rho} - 2\, \eta_{\mu \rho} \, \eta_{\nu \sigma} \, , &
\tr \left(P_\mu \, K_\nu \right) &= 4 \, \eta_{\mu \nu} \, , & 
\tr \left( D \, D \right) &= 2 \, ,	
\label{eqn:Metric}
\end{align}
and all other components vanish. The $\mathbb{Z}_2$ grading of the algebra gives the decomposition 
\begin{align}
\alg{h} &= \mathrm{span} \left \lbrace M_{\mu \nu} , P_\mu - K_\mu \right \rbrace  \, , &
\alg{m} &= \mathrm{span} \left \lbrace P_\mu + K_\mu , D \right \rbrace \, ,
\end{align}
and it is easy to show that the Lie algebra $\alg{h}$ of the gauge group is indeed isomorphic to 
$\alg{so}(1,4)$. 

In order to introduce coordinates in this formalism, we choose a set of coset representatives. To obtain 
Poincar{\'e} coordinates, which are appropriate for the study of minimal surfaces, the 
following coset representatives are a good choice:
\begin{align}
g(X,y) = e^{X \cdot P} \, y^D . 
\end{align}
The Maurer--Cartan form $U$ is then given by 
\begin{align}
U = g^{-1} \diff g = \frac{\diff X^\mu}{y} \, P_\mu + \frac{\diff y}{y} \, D ,  
\label{eqn:UXy}
\end{align}
which the reader is invited to check. For the projections, we note
\begin{align}
A &=  \frac{\diff X^\mu}{2y} \left( P_\mu - K_\mu \right) \, ,  &
a &= \frac{\diff X^\mu}{2y} \left( P_\mu + K_\mu \right)+ \frac{\diff y}{y} \, D .
\end{align}

The metric for the coset space is then obtained as
\begin{align}
\Gamma_{ij} = \tr \left( a _i \, a_j \right) 
	= \frac{ \eta^{\mu \nu}  \partial_i X_\mu  \, \partial_j X_\nu 
	+ \partial _i y \, \partial_j y }{y^2} ,
\end{align}
showing that our coset representatives indeed correspond to Poincar{\'e} coordinates on $\AdS_5$.  

\subsection{Conserved Charges}
Due to the integrability of symmetric space models, we can construct an infinite set of 
conserved charges. The construction is based on the Lax connection $L_u$, which is a 
one-parameter family of flat connections, i.e.\ for every value of the spectral parameter 
$u$, we have 
\begin{align}
\partial_\tau L_{u, \sigma} - \partial_\sigma L_{u, \tau} 
	+ \brk[s]*{L_{u, \tau} , L_{u, \sigma} } = 0 . 
\end{align}
The flatness of the connection implies that --- at least locally --- the auxiliary 
linear problem
\begin{align}
\partial_\tau \Psi &= \Psi L_{u, \tau}  \, , & 
\partial_\sigma \Psi &= \Psi L_{u, \sigma}  
\end{align}
has a solution, since the two conditions are compatible. Note that the Maurer--Cartan form 
$U$ is flat by construction. In the present case, we can construct a Lax connection from the 
components of $U$ by setting (we employ conformal gauge)
\begin{align}
L_{u, \tau} &= A_\tau + \frac{1-u^2}{1+u^2} a_\tau 
	+ \frac{2u}{1+u^2} a_\sigma \, , \\ 
L_{u, \sigma} &= A_\sigma + \frac{1-u^2}{1+u^2} a_\sigma 
	- \frac{2u}{1+u^2} a_\tau . 
\end{align}
When checking the flatness of the above connection or performing similar 
calculations, the reader is advised to make use of the language of differential forms, 
in which we may write the Lax connection more compactly as 
\begin{align}
L_u = A + \frac{1-u^2}{1+u^2} a - \frac{2u}{1+u^2} \ast a
\end{align} 
The flatness condition $\diff L_u + L_u \wedge L_u = 0$ can be derived straightforwardly%
\footnote{For the Hodge-star operator on the worldsheet, we note the helpful identities 
\begin{align*}
\ast (\ast r ) &= - r \, , & 
\ast r \wedge s &= - r \wedge \ast s  \, ,
\end{align*}
which hold for generic one-forms $r$ and $s$ and the first identity requires the 
worldsheet metric to have Euclidean signature, otherwise there is a sign flip.}
from the equations of motion 
\begin{align}
\diff \ast a + A \wedge \ast a + \ast a \wedge A = 0 \, .
\end{align}

The conserved charges are obtained from the monodromy over the Lax connection, 
\begin{align}
T_u = \prexp \brk*{ \int \diff \sigma \, L_\sigma } .  
\end{align}
Here, we integrate over slices of constant $\tau$. The $\tau$-dependence of the monodromy is 
described by the evolution equation 
\begin{align}
\partial _\tau T_u = \brk[s]*{ T_u , L_{u, \tau} (\tau, \sigma = 0) } ,  
\end{align}
which the reader may derive by noting that $T_u$ is indeed a monodromy for the auxiliary 
linear problem, i.e.\ it satisfies 
\begin{align}
\Psi(\tau, 2 \pi ) = \Psi(\tau, 0 ) T_u (\tau) . 
\end{align}
A reformulation of the equations of motion in this way is known as a Lax pair. It implies that 
the eigenvalues of the monodromy are conserved quantities. 
One way to see this is to show that the evolution equation can be solved by considering $\tau$-dependent 
similarity transformations of $T_u (\tau_0)$, 
\begin{align*}
T_u (\tau) = S_u (\tau) \, T_u (\tau_0) \, S_u (\tau)^{-1} .
\end{align*}
Plugging this into the evolution equation for the monodromy leads to an equation for the transformation 
matrix $S_u (\tau)$, for which a solution exists. 

For minimal surfaces, the situation is special. Note that the minimal surface closes and thus has 
the topology of a disc. This means that we can contract any curve on the minimal surface to a point. 
In the case at hand this tells us that the monodromy will become trivial for some value of $\tau$, 
i.e.\ we have 
\begin{align}
T_u (\tau_0) = \unit , 
\end{align}
and by similarity, this extends to all values of $\tau$. The monodromy is thus indeed conserved as 
a whole. 

The fact that the monodromy is trivial exhibits global information about the minimal surface, 
it does not follow from the equations of motion and can be read as a constraint on the 
unfixed coefficients in the Polyakov--Rychkov expansion --- they need to be adjusted in such 
a way that the minimal surface closes. These constraints are, however, difficult to extract 
from the monodromy in the form we have given above. 
In order to reach a more useful form, we employ a flatness-preserving transformation of the form
\begin{align}
L_u \; \mapsto \; L_u ^\prime = f^{-1} L_u f + f^{-1} \diff f \, , 
\end{align}
In our case, we specifically consider the case $f = g^{-1}$ to reach the transformed connection
\begin{align}
L_u ^\prime = \ell_u = g L_u g^{-1} - \diff g \, g^{-1} 
	= g (L_u - U) g^{-1} 
	= \frac{1}{1+u^2} \brk*{u \, \ast j - u^2 j } .
\end{align}
Here, $j = - 2 g a g^{-1}$ denotes the Noether current of the model, or rather a collection 
of all the Noether currents associated to the G-symmetries in a single matrix. 
Under the above transformation, the monodromy transforms by a similarity transformation, 
\begin{align}
T_u &= g_0 \,  t_u \,  g_0^{-1} \, , & 
t_u &= \prexp \left( \int \diff \sigma \, \ell_{u, \sigma} \right)  .  
\end{align}
In order to prove this transformation behavior, the reader may rely on the same techniques we 
employed to derive similar identities for the Wilson loop. 

We can now extract conserved charges from the expansion of the monodromy $t_u$ around $u=0$, 
\begin{align}
t_u = \exp \brk*{ u \, Q^{(0)} + \half u^2 \, Q^{(1)} + \ldots } .
\end{align}
Note that $t_u$ takes values in the Lie group G and we have organized the expansion in such 
a way that the charges take values in the Lie algebra $\alg{g}$. 
Other charges can be obtained from expanding around different points, but that is not our concern 
here. The Lax connection $\ell_u$ has the expansion 
\begin{align*}
\ell_u = u \ast j - u^2 j + \Order (u^3) ,  
\end{align*} 
and we read off the conserved charges 
\begin{align}
Q^{(0)} &= - \int \limits _0 ^L \diff \sigma \, j_\tau \, , \\
Q^{(1)} &=  \int \limits _{0} ^L \diff \sigma_1 \, \diff \sigma_2 \, 
	\theta \left( \sigma_2 - \sigma_1 \right) 
	\left[ j_\tau (\sigma_1) , j_\tau (\sigma_2) \right]
	- 2 \int \limits _{0} ^L \diff \sigma \, j_\sigma (\sigma) .
\end{align}
The Poisson algebra of these charges forms the classical counterpart of 
a Yangian algebra \cite{MacKay:1992he,Klose:2016uur}. Related algebraic 
structures are discussed in the review on One-point functions in AdS/dCFT
\cite{deLeeuw:2019usb} to appear in the same special issue of J.\ Phys.\ A.
For an introduction to Yangian symmmetry, the reader is invited to consult references 
\cite{MacKay:2004tc,Torrielli:2011gg,Ferro:2011ph,Loebbert:2016cdm}. 
Below, we will extract Yangian symmetry generators for the Maldacena--Wilson loop at strong coupling 
from the finding that these charges vanish, which follows directly from the triviality of the monodromy. 

\subsection{Yangian Symmetry for Minimal Surfaces}

In order to do so, we return to the minimal surfaces in $\AdS_5$ and evaluate the charges 
on the minimal surface by making use of the Polyakov--Rychkov expansion
\eqref{eqn:expansion_boundary_arc}. The relevant information to be obtained from the 
$\tau$-expansion of the conserved charges is the vanishing of the $\tau^0$-coefficient, which contains
the global information about the minimal surface, whereas the vanishing of the other coefficients 
follows directly from the conservation of the charges, i.e.\ from the equations of motion. 

We are thus interested in the $\tau^0$-coefficient of the Noether current $j_\tau$, 
\begin{align}
j_\tau = -2g a_\tau g^{-1} =
	- 2 \, \eunit ^{XP} \left( 
	\frac{\partial_\tau X^\mu}{2 y^2} (K_\mu + y^2 P_\mu) 
	+ \frac{\partial_\tau y}{y} D \right) \eunit ^{-XP} . 
\end{align}
Inserting the expansion \eqref{eqn:expansion_boundary}, we find that
\begin{align*}
j_\tau = - \, \eunit ^{XP} \left( 
	\frac{\ddx_\mu}{\tau} K^\mu + \frac{2}{\tau} D 
	- \frac{\delta A_{\mathrm{ren}}(\gamma)}{\delta x^\mu (\sigma)} K^\mu 
	+ \Order ( \tau )
	\right) \eunit ^{-XP} .
\end{align*}
Note now that, since $X = x + \Order ( \tau^2 )$, the conjugation with $\eunit ^{XP}$ 
does not mix the $\tau^{-1}$-order and the $\tau^0$-order, such that the $\tau^0$-coefficient 
is found to be given by
\begin{align}
j_{\tau \, (0)} &= 4 \, \frac{\delta A_{\mathrm{ren}}(\gamma)}
	{\delta x^\mu}  \, \hat{\xi}^\mu (x) \, , 
\label{eqn:jtau0}	
\end{align}
where $\hat{\xi}^\mu (x)$ comprises the conformal Killing vectors 
\begin{align}
\hat{\xi}^\mu (x) & = \xi ^\mu _a (x) \, T^a = \frac{1}{4} \, e^{x P} K^\mu e^{-x P} \, , \\ 
	\xi^\mu _a (x) &= \left \lbrace  \delta ^\mu _\nu , \,
	 x_\nu \delta ^\mu _\rho - x_\rho \delta ^\mu _\nu , \, 
	 x^\mu , \,
	x^2 \delta ^\mu _\nu - 2 x^\mu x_\nu \right \rbrace . 
\end{align}
The appearance of the conformal Killing vectors is not surprising: In general, the Noether current 
contains the Killing vectors of the underlying space and in the limit toward the conformal boundary 
we obtain the conformal Killing vectors of the boundary space. 

The vanishing of the charge $Q^{(0)}$ thus entails the conformal symmetry of the minimal area, 
\begin{align}
\int \diff \sigma \, \xi^\mu _a (x) \frac{\delta A_{\mathrm{ren}}(\gamma)}
	{\delta x^\mu (\sigma)}
	= 0 .
\end{align}
The vanishing of the charge $Q^{(1)}$ leads to a more interesting symmetry. After some 
calculation, we obtain the identity  
\begin{align}
\mathbf{f} \indices{_a ^{cb}} \, \int \limits _{0} ^L \diff \sigma_1 \, \diff \sigma_2 \, 
	\theta (\sigma_2 - \sigma_1) \,  \xi ^\mu _{1 b} \, 
	\frac{\delta A_\mathrm{ren}}{\delta x_1^\mu  } \, 
	\xi ^\nu _{2 c} \, 
	\frac{\delta A_\mathrm{ren}}{\delta x_2^\nu } 
	- \frac{1}{2}  \int \limits _0 ^L \diff \sigma \, \xi ^\mu _a
	\left( \dx_\mu \, \ddot{x}^2 + \dddot{x}_\mu \right) = 0 . 
\label{Q1identity}	
\end{align}
Here, $\mathbf{f} \indices{_a ^{cb}}$ are the (dual) structure constants of the conformal 
algebra, which follow e.g.\ from the Lie bracket of the conformal Killing vector fields, 
\begin{align}
\left \lbrace \xi _a , \xi _b \right \rbrace ^\mu 
	= \xi ^\nu _a \partial_\nu \, \xi ^\mu _b 
	- \xi ^\nu _b \partial_\nu \, \xi ^\mu _a 
	= \mathbf{f} _{ab} {} ^c \, \xi ^\mu _c .
\end{align}
One way to interpret this identity is to note that it arises from the application of the 
generator 
\begin{align}
\mathrm{J}_a ^{(1)} = \mathbf{f} \indices{_a ^{cb}} \, \int \limits _{0} ^L 
	\diff \sigma_1 \, \diff \sigma_2 \, 
	\theta (\sigma_2 - \sigma_1) \,  \xi ^\mu _{1 b} \,  \xi ^\nu _{2 c} \,  
	\frac{\delta^2}{\delta x^\mu _1 \delta x^\nu _2} 
	- \frac{\la}{8 \pi ^2} \int \limits _0 ^L \diff \sigma \, \xi^\mu _a  
	\left( \dx_\mu \, \ddot{x}^2 + \dddot{x}_\mu \right) 
\label{J1Bos}	
\end{align}
to the expectation value of the Maldacena--Wilson loop at strong coupling, 
\begin{align}
\left \langle W(\gamma) \right \rangle =
	\exp \left( - \ft{\sqrt{\lambda}}{2 \pi} A_\mathrm{ren}(\gamma) \right) .
\end{align}
Indeed the generator $\mathrm{J}_a ^{(1)}$ has the typical form of a level-1 Yangian symmetry 
generator and satisfies the respective algebra. This finding is naturally related to the finding 
that the Poisson algebra of the conserved charges is the classical counterpart of a Yangian algebra. 

Demanding that $\mathrm{J}_a ^{(1)} \left \langle W(\gamma) \right \rangle = 0$ 
gives the identity \eqref{Q1identity} at the leading order in $\lambda$. 
The application of the same generator to the expectation value \eqref{W1loop} of the 
Maldacena--Wilson loop at weak coupling shows, however, that it is not a symmetry there
\cite{Muller:2013rta}.  This finding can be understood from the fact that the generator 
$\mathrm{J}_a ^{(1)}$ fails to be cyclic. 

One way to obtain cyclic generators is to consider an underlying Lie algebra for which 
the contraction $\mathbf{f} \indices{_a ^{cb}} \mathbf{f} \indices{_{bc}^d}$ vanishes, which 
is for example the case for the superconformal algebra $\alg{psu}(2,2 \vert 4)$. This is indeed 
the symmetry algebra for the Wilson loops in superspace, which generalize the Maldacena--Wilson 
loop to a non-chiral $\mathcal{N}\!=4$ superspace. These loop operators have been constructed and shown 
to be Yangian symmetric both at weak and strong coupling  
\cite{Beisert:2015jxa,Beisert:2015uda,Munkler:2015gja}. 

\subsection{Spectral-parameter Deformation or Master Symmetry}

\begin{figure}
\centering
\includegraphics[width=75mm]{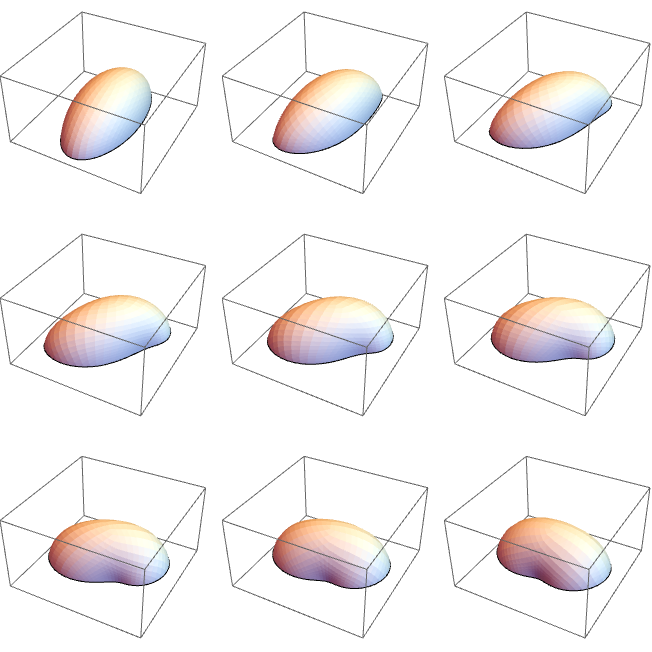}
\caption{Minimal surfaces arising from the spectral-parameter deformation of the minimal surface
for an elliptical boundary curve. The values of the spectral parameter $\theta$ range from $0$ 
to $\pi$ in uniform steps. For values ranging from $\pi$ to $2\pi$, the deformation continues 
until the original shape is reached again. 
The figure is based on numerical data and has been reproduced 
from reference \cite{Klose:2016qfv}.}
\label{fig:ellipse-family}
\end{figure}

There is another symmetry of minimal surfaces in $\mathrm{AdS}_5$, which is given by a 
one-parameter group of deformations of the minimal surface known as spectral-parameter
deformations \cite{Ishizeki:2011bf,Kruczenski:2013bsa,Kruczenski:2014bla,Dekel:2015bla}, 
see figure \ref{fig:ellipse-family} for a depiction of the deformations of an ellipse 
based on a numerical evaluation \cite{Klose:2016qfv}. All of the surfaces shown there 
have the same area. 

This symmetry can be reduced to the finding that upon replacing 
\begin{align}
U \; \; \mapsto \; \; L_u 
\end{align}
the action is invariant and the equations of motion are still satisfied. For the action, 
this follows immediately by replacing
\begin{align}
a \; \; \mapsto \; \; a_u = \frac{1-u^2}{1+u^2} \, a
	- \frac{2u}{1+u^2} \, \ast a \, ,
\end{align}
which implies that
\begin{align}
A_{P, u} = \frac{1}{2} \int \tr ( a_u \wedge \ast a_u ) 
	= \frac{1}{2}\, \frac{(1-u^2)^2 + 4 u^2}{(1+u^2)^2} 
	\int  \tr \left( a \wedge \ast a \right) = A_P .
\end{align}
In a similar fashion, one may show that $L_u$ provides a solution of the equations of motion 
if $U$ does.

In order to transfer the symmetry to the fields $g(\tau , \sigma)$, we require that the 
deformed solution $M_u[g]$ has the Lax connection $L_u$ as its Maurer--Cartan form, 
i.e.\ we demand that 
\begin{align} 
M_u[g] ^{-1} \diff M_u[g] &= L_u [g] \, , &
M_u[g] (\tau_0,\sigma_0) &= g (\tau_0,\sigma_0) \, ,
\label{eqn:def-deformed-sol}
\end{align}
The deformation $M_u[g]$ is well-defined if the Lax connection $L_u$ is flat, 
i.e.\ when $g(\tau , \sigma)$ is a solution of the equations of motion. 

Let us now work out how this symmetry is related to the ones we have discussed above. 
The relation to the conformal transformations or $\mathrm{AdS}$-isometries is easy to establish. 
In the coset-description, these symmetries are realized by left-multiplication with a constant 
group element, $g \to L \cdot g$. 
Using that the solution to equation \eqref{eqn:def-deformed-sol} is unique, we can then show 
that $M_u[L \cdot g] = L \cdot M_u[g]$. This follows directly by plugging it into the defining 
equation and using that $(L \cdot g)^{-1} \diff (L \cdot g) = g^{-1} \diff g$ for 
$L \in \grp{G}$ constant. 
The spectral-parameter deformations thus commute with the conformal transformations. 
A concatenation of two spectral-parameter deformations can be worked out in the same way 
and results in the identity  
\begin{align}
M_{u_1} [ M_{u_2} [ g]  ] = M_{(u_1+u_2)/(1+u_1u_2)}[g]
\end{align}

The relation to the Yangian symmetries we have discussed above is more difficult to establish. 
We begin by considering the variation $\master$ associated to the spectral-parameter 
transformation, which is given by
\begin{align}
\master g = \frac{\diff}{\diff u} M_u[g] \vert _{u=0} &= \chi^{(0)} \cdot g \, , & 
\chi^{(0)} (\tau , \sigma) = \int \limits _{(\tau_0 , \sigma_0)} ^{(\tau , \sigma)} \ast j .
\label{eqn:master-var}
\end{align}
The variation associated to the Yangian symmetry is given by \cite{Dolan:1980kz}
\begin{align}
\delta _\epsilon ^{(1)} g = \left[ \chi^{(0)} , \epsilon \right] g . 
\end{align}
In fact, it is part of an infinite tower of symmetry variations $\delta _\epsilon ^{(n)}$, which 
begins with the variation 
 \begin{align}
\delta _\epsilon ^{(0)} g = \delta _\epsilon ^{(0)} g 
	= \epsilon \cdot g  
\end{align}
associated to the $\mathrm{AdS}$-isometries. The higher-order variations contain $n$-point 
integrals of the form \eqref{eqn:master-var} and are related to the higher-level generators of 
the Yangian symmetry. Using this set-up, we can discuss the relation between the 
spectral-parameter deformation and the Yangian symmetries by calculating the commutation 
relations between these variations. This gives 
\begin{align}
\left[ \master , \delta _\epsilon ^{(1)} \right] = \delta _\epsilon ^{(2)} 
	- \delta_{\epsilon'} ^{(0)} \, ,  
\end{align} 
showing that the spectral-parameter deformations can be employed to construct the higher-level 
symmetries of the Yangian. For this reason, the deformation has been called the master 
symmetry in reference \cite{Klose:2016qfv}, where the reader can find a much more detailed 
description of their algebraic properties. 

\section*{Acknowledgements}
This review is based on lectures given at the Young Researchers Integrability School 
and Workshop in Ascona in 2018. I am happy to thank the organizers and the scientific committee
of this school for their organizational efforts and for the invitation to provide a lecture on 
Wilson loops. I am thankful to the participants for their questions and helpful remarks during 
our discussions. I would also like to thank the other lecturers in Ascona for the inspiring exchange and
fruitful coordination of the different courses. Let me also thank Daniel Medina Rincon and 
Tristan McLoughlin for their valuable comments on the draft. 

This work has been supported by the grant no.\ 615203 from the European Research Council under the FP7 and by the 
Swiss National Science Foundation through the NCCR SwissMAP. 

\bibliographystyle{nb}
\bibliography{lecture-notes}

\end{document}